\documentclass[sigconf, nonacm]{acmart}
\usepackage{tikz}
\usepackage{booktabs}
\usepackage{multirow}
\usepackage{subcaption}
\usepackage{graphicx}
\usepackage{tcolorbox}
\usepackage{float}
\usepackage{siunitx}
\usepackage[utf8]{inputenc}
\usepackage{tabularx}
\usepackage{amsmath}
\usepackage{enumitem}

\usepackage{tikz}
\usetikzlibrary{positioning, arrows.meta, calc, chains, decorations.pathreplacing}
\usepackage{pgfplots}
\pgfplotsset{compat=1.18}
\colorlet{stStorage}{black!8}
\colorlet{stDecode}{black!45}
\colorlet{stPre}{black!25}
\colorlet{stXfer}{black!8}
\colorlet{stCompute}{black!65}
\tikzset{stXferPat/.style={postaction={pattern=dots, pattern color=black!30}}}
\newcommand{\stagetag}[3][]{\tikz[baseline=(t.base)]\node[inner sep=1.6pt, minimum width=1.1em, fill=#2, draw=black!50, line width=0.3pt, font=\scriptsize\sffamily\bfseries, #1](t){#3};}
\usepackage{calc}
\usepackage{xcolor}
\usepackage{caption}

\usepackage{graphicx}
\usepackage{array}
\usepackage{booktabs}
\usepackage{multirow}

\tcbuselibrary{skins}
\newcounter{insight}
\newtcolorbox{insight}[1][]{%
  enhanced, colback=black!4, colframe=black!4, boxrule=0pt, arc=0pt,
  borderline west={1.6pt}{0pt}{black!75},
  left=5pt, right=4pt, top=2.5pt, bottom=2.5pt,
  before skip=6pt, after skip=6pt, fontupper=\small\itshape,
  before upper={\refstepcounter{insight}{\upshape\bfseries Insight~\theinsight.}\ },
  #1}

\newtcolorbox{rqbox}{%
  enhanced, colback=white, colframe=black, boxrule=0.4pt, arc=0pt,
  left=4pt, right=4pt, top=2pt, bottom=2pt, before skip=5pt, after skip=5pt}

\newlength{\bnumcap}
\newcommand*{\bnum}[1]{\settoheight{\bnumcap}{H}%
  \tikz[baseline={([yshift=-0.5\bnumcap]c.center)}]\node[circle,draw,line width=0.7pt,inner sep=0pt,minimum size=0.8em,font=\sffamily\bfseries\footnotesize](c){#1};}

\usepackage{listings}
\definecolor{codegreen}{rgb}{0.13, 0.55, 0.13}
\definecolor{codeblue}{rgb}{0.0, 0.2, 0.6}
\definecolor{codegray}{rgb}{0.5, 0.5, 0.5}
\definecolor{codepurple}{rgb}{0.58, 0, 0.82}
\definecolor{backcolour}{rgb}{0.97, 0.97, 0.97}
\lstdefinestyle{vldbstyle}{
    backgroundcolor=\color{backcolour},   
    commentstyle=\color{codegreen}\itshape,
    keywordstyle=\color{codeblue}\bfseries,
    numberstyle=\tiny\color{codegray},
    stringstyle=\color{codepurple},
    basicstyle=\ttfamily\footnotesize,
    breakatwhitespace=false,         
    breaklines=true,                 
    captionpos=b,                    
    keepspaces=true,                 
    numbers=left,
    numbersep=5pt,                  
    showspaces=false,                
    showstringspaces=false,
    showtabs=false,                  
    tabsize=4,
    frame=lines,
    rulecolor=\color{black!30},
    xleftmargin=1em,
    framexleftmargin=1em
}

\usepackage[capitalise,nameinlink]{cleveref} 

\usetikzlibrary{patterns}

\newcommand{\boldfirst}[1]{\noindent\textbf{#1}\hspace{0.5em plus 0.2em}\ignorespaces}
\begin{document}
\title{Anatomy of a Data Pipeline in Deep Learning Training [Experiment, Analysis \& Benchmark]}
\title{Anatomy of a Data Pipeline in Deep Learning Training}

\author{Jens Birk Andersen}
\affiliation{%
  \institution{Samsung}
}
\affiliation{%
  \institution{IT University of Copenhagen}
}
\email{jb.andersen@samsung.com}

\author{Vivek Shah}
\author{Javier González}
\affiliation{%
  \institution{Samsung}
}
\email{vi.shah@samsung.com}
\email{javier.gonz@samsung.com}

\author{Pınar Tözün}
\affiliation{%
  \institution{IT University of Copenhagen}
}
\email{pito@itu.dk}

\begin{abstract}

Modern Deep Learning (DL) training runs on deeply layered software stacks that make the journey of data from storage to an accelerator hard to understand. Performance problems are therefore often attributed to data movement, and mitigation strategies are designed for it without analyzing where time actually goes.
To better characterize the end-to-end performance and resource utilization of diverse modern DL pipelines, we study these pipelines across four domains of Natural Language Processing (NLP), Computer Vision, Medical Imaging, and Speech, using 11 open-weight models of varying size and type. We base our study on widely used open-source frameworks centered on PyTorch as the runtime, using high-performance NVIDIA H100 GPUs and NVMe SSDs.

Our findings show that storage is not the bottleneck. It accounts for at most 7\% of per-batch latency, and raw NVMe throughput exceeds end-to-end training throughput by 30--203$\times$ outside NLP and orders of magnitude in NLP. Bottlenecks instead depend on the domain. NLP training is compute-bound with Llama-3.2-3B reading under 50~KiB/s from disk while keeping an H100 busy. By contrast, in Medical Imaging, decoding takes up to 83\% of per-batch latency. We also find inefficiencies across the open-source deep learning software ecosystem. For example, multiprocessing dataloaders grow host (CPU) memory by 137~GiB at 48 workers and epoch boundaries stall training for up to 65 seconds. 
Furthermore, on NVMe SSDs, small checkpoints such as SegResNet's and 3D-UNet's (0.05~GiB) can be written every step at under 9\% throughput cost, while Llama-3.2-3B's 18~GiB checkpoint written every 100 steps keeps 89\% of throughput asynchronously but only 70\% synchronously. We distill these results into domain-specific guidance for deploying and improving resource-efficient DL systems.

\end{abstract}

\maketitle

\section{Introduction}
\label{sec:intro}
Deep Learning (DL) model sizes and training datasets have grown by orders of magnitude over the past decade, enabling breakthroughs in large language models~\cite{brown_language_2020}, vision models~\cite{dosovitskiy_image_2021, he_deep_2015}, and multimodal systems~\cite{radford_learning_2021}. Sustaining this progress requires every component of the training pipeline to scale alongside the models.

While much effort has gone into accelerating model computation through more powerful GPUs~\cite{noauthor_nvidia_nodate}, distributed training algorithms, and high-performance interconnects~\cite{rajbhandari_zero-infinity_2021}, the data pipeline matters just as much for end-to-end training performance~\cite{mohan_analyzing_2021, isenko_where_2022, cedarVLDB24, murray_tfdata_2021, metaDataLoadingAnalysis, kuchnik_plumber_2022}. Training jobs repeatedly read massive datasets from storage, decode and preprocess samples, and deliver them to accelerators fast enough to avoid GPU idle time. As datasets grow, inefficiencies in data loading can limit hardware utilization and increase training time and cost.

However,
existing work provides only fragmented views of the training pipeline. Storage-oriented benchmarks such as MLPerf Storage \cite{Balmau22} and DLIO~\cite{devarajan_dlio_2021} accurately reproduce storage access patterns, but replace GPU execution with synthetic computation, which hides insights into the data rate needed to keep the GPU busy. Conversely, MLPerf Training~\cite{mattson_mlperf_2020} measures end-to-end training performance, but its focus on time-to-accuracy does not show which pipeline stage limits training. Earlier bottleneck studies introduced the concept of data stalls~\cite{isenko_where_2022, mohan_analyzing_2021}, establishing that both ``fetch'' and ``prep'' stalls were frequently the limiting factor in DNN training. These studies, however, were mostly conducted on CNN-era vision workloads (e.g., ResNet on ImageNet) running on GPUs roughly an order of magnitude less powerful than today's~\cite{nvidia_v100_2020, noauthor_nvidia_nodate}\footnote{125 tensor Tflop/s (V100) vs.\ 989--1,979 dense Tflop/s (H100 SXM, FP16--FP8).}.

Two things have changed since. First, accelerator throughput has scaled much faster than CPU performance~\cite{HennessyP19} and memory and interconnect bandwidth~\cite{ai_memory_wall_24}, tightening the compute budget available for masking preprocessing costs. Second, the workload mix has shifted toward Transformer-based architectures whose arithmetic intensity (flop per byte of HBM traffic) is substantially higher than that of CNNs~\cite{wang_benchmarking_2019}. Whether the original data-stall diagnosis still holds, and for which domains, has not been re-examined on current hardware and models. As a result, practitioners still lack a detailed end-to-end characterization of where time is spent across the modern deep learning data path.

This paper characterizes, stage by stage, where modern DL training workloads are bottlenecked with a specific focus on the data path.\footnote{Code, scripts, and data are available at
\url{https://github.com/jensbirk/anatomy-data-pipeline}.} We trace training samples from persistent storage through format decoding, preprocessing, host-to-device transfer, GPU execution, and checkpointing, spanning four application domains, eleven representative models, multiple storage formats, and four widely used dataloading frameworks. Our analysis shows that storage I/O is not the dominant cause of data stalls. Modern NVMe storage delivers 30--203$\times$ more throughput than end-to-end training requires outside NLP, and five orders of magnitude more for NLP, where Llama-3.2-3B reads under 50 KiB/s while keeping an H100 busy. Instead, bottlenecks shift across domains. NLP workloads are GPU compute-bound, whereas vision and especially medical imaging pipelines are dominated by CPU-side decoding and preprocessing. Furthermore, NLP and Speech train at full throughput under host-memory limits of only 1 and 5 GiB while PyTorch's DataLoader grows host memory by 137 GiB at 48 workers although throughput saturates at 8 for Vision. Lastly, epoch boundaries stall training for up to 65 s, and checkpointing Llama-3.2-3B every 100 steps writes 18--23 TB per day, two to three full writes of a 7.68 TB SSD.

Our contributions are:
\begin{list}{\labelitemi}{\leftmargin=1.5em}
\item{A detailed decomposition of the modern model training data path and bottlenecks into software and hardware layers from persistent storage to GPU execution (\autoref{sec:background}).}
\item{An end-to-end characterization of the data path across four application domains, eleven models, multiple storage formats, and four dataloading frameworks, leading to new empirical insights, especially for large-language model training (\autoref{sec:data_loading}).}
\item{A characterization of checkpointing performance, quantifying the impact of checkpoint formats, checkpoint frequency, and synchronous and asynchronous checkpointing on local and remote storage (\autoref{sec:checkpointing}).}
\item{Practical guidance on hardware configurations and software setups based on our results (\autoref{sec:discussion}).}
\end{list}

Additionally,
\autoref{sec:methodology} details our experimental methodology, 
\autoref{sec:related} surveys related work,
and \autoref{sec:conclusion} concludes.

\section{DATA PIPELINES UNDER THE MICROSCOPE}
\label{sec:background}
Analyzing bottlenecks requires understanding the computational and I/O phases of deep learning training. To isolate where and why these bottlenecks occur, we trace the end-to-end data pipeline using a top-down approach. First, we establish the distinct modalities, algorithmic requirements, and preprocessing steps across different deep learning domains (\autoref{sec:background:pipelines}). We then define data pipelines in DL training (\autoref{sec:background:data_pipelines}). Next, we examine the software stack and architectures responsible for orchestrating data movement (\autoref{sec:background:software_stack}). Finally, we map these software demands onto the underlying hardware and classify where data stalls can arise (\autoref{sec:background:performance_hardware}) and describe the checkpoint write path (\autoref{sec:background:checkpointing}).

\subsection{Domain-Specific Data Processing}
\label{sec:background:pipelines}

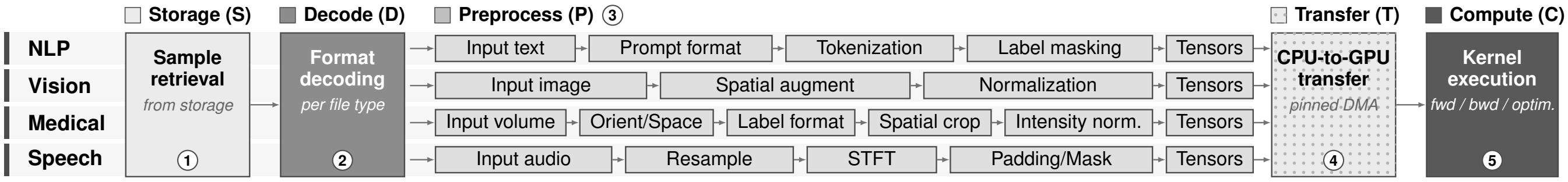
\begin{figure*}[t!]    
    \centering
    {\fontfamily{phv}\selectfont
\definecolor{cG}{HTML}{4D4D4D}

\begin{tikzpicture}[x=1pt,y=1pt,font=\fontsize{6.2}{7}\selectfont,
  chip/.style={draw=cG, sharp corners, line width=0.45pt,
               inner xsep=2pt, inner ysep=1pt, minimum height=8pt,
               fill=cG!18, text=black, anchor=west, text height=5pt, text depth=1.4pt},
  lbl/.style={anchor=west, font=\fontsize{6.8}{7.5}\selectfont\bfseries},
  hdr/.style={anchor=west, font=\fontsize{6.2}{7}\selectfont\bfseries,
              text height=5pt, text depth=1.4pt, inner xsep=1.5pt},
  tall/.style={draw=black!70, line width=0.45pt, sharp corners, align=center,
               minimum width=40pt, minimum height=46.5pt, inner sep=1pt,
               anchor=north west},
  sub/.style={font=\fontsize{5.2}{6}\selectfont\itshape},
  flow/.style={-{Latex[length=2.4pt]}, black!55, line width=0.45pt},
  lst/.style={circle, draw=black!70, line width=0.4pt, inner sep=0.3pt,
              minimum size=6.4pt, font=\fontsize{4.8}{5}\selectfont\bfseries,
              fill=white, text=black},
]

\newcommand{\figonehd}[4]{%
  \fill[#2] (#1,48.5) rectangle +(5,5);
  \draw[black!70, line width=0.3pt] (#1,48.5) rectangle +(5,5);
  \node[hdr] (#4) at (#1+6,51) {#3};
}
\figonehd{36}{stStorage}{Storage (S)}{hS}
\figonehd{86}{stDecode}{Decode (D)}{hD}
\figonehd{136}{stPre}{Preprocess (P)}{hP}
  \node[lst, right=1pt of hP] {3};
\figonehd{406}{stXfer}{Transfer (T)}{hT}
\path[stXferPat] (406,48.5) rectangle +(5,5);
\figonehd{456}{stCompute}{Compute (C)}{hC}

\newcommand{\figonerow}[2]{%
  \fill[black!3.5] (-3,#1) rectangle (400,#1-10.5);
  \fill[cG] (-3,#1) rectangle (-1.4,#1-10.5);
  \node[lbl] at (1.5,#1-5.25) {#2};
}
\figonerow{45.5}{NLP}
\figonerow{33.5}{Vision}
\figonerow{21.5}{Medical}
\figonerow{9.5}{Speech}

\node[tall, fill=stStorage] (S) at (36,45.5)
  {\textbf{Sample}\\\textbf{retrieval}\\[1pt]{\fontsize{5.2}{6}\selectfont\itshape\color{black!60}from storage}\\[9pt]};
\node[lst] at ([yshift=5.5pt]S.south) {1};

\node[tall, fill=stDecode, text=white] (D) at (86,45.5)
  {\textbf{Format}\\\textbf{decoding}\\[1pt]{\fontsize{5.2}{6}\selectfont\itshape per file type}\\[9pt]};
\node[lst] at ([yshift=5.5pt]D.south) {2};

\node[tall, fill=stXfer, stXferPat] (T) at (406,45.5)
  {\textbf{CPU-to-GPU}\\\textbf{transfer}\\[1pt]{\fontsize{5.2}{6}\selectfont\itshape\color{black!60}pinned DMA}\\[9pt]};
\node[lst] at ([yshift=5.5pt]T.south) {4};

\node[tall, fill=stCompute, text=white] (C) at (456,45.5)
  {\textbf{Kernel}\\\textbf{execution}\\[1pt]{\fontsize{5.2}{6}\selectfont\itshape fwd / bwd / optim.}\\[9pt]};
\node[lst] at ([yshift=5.5pt]C.south) {5};

\draw[flow] (S.east) -- (D.west);
\draw[flow] (T.east) -- (C.west);

\def\Px{136}      
\def\Tx{372}      
\def\Tw{28}       
\def\gap{4.5}     

\newcommand{\figonetxtw}[1]{\begingroup\fontsize{6.2}{7}\selectfont
  \pgfmathwidth{"#1"}\xdef\tw{\pgfmathresult}\endgroup}

\newcommand{\figonepipe}[3]{%
  
  \xdef\sumw{0}\xdef\nch{0}%
  \foreach \p in {#3} {%
    \figonetxtw{\p}%
    \pgfmathsetmacro{\a}{\sumw+\tw}\xdef\sumw{\a}%
    \pgfmathtruncatemacro{\b}{\nch+1}\xdef\nch{\b}}%
  
  \pgfmathsetmacro{\ex}{((\Tx-\gap)-\Px-\sumw-4*\nch-(\nch-1)*\gap)/\nch}%
  \xdef\ex{\ex}\xdef\cx{\Px}%
  
  \foreach \p [count=\i] in {#3} {%
    \figonetxtw{\p}%
    \pgfmathsetmacro{\w}{\tw+4+\ex}%
    \node[chip, minimum width=\w pt] (#1p\i) at (\cx,#2) {\p};%
    \ifnum\i=1 \draw[flow] (128,#2) -- (#1p\i.west);%
    \else \pgfmathtruncatemacro{\j}{\i-1}\draw[flow] (#1p\j.east) -- (#1p\i.west);\fi
    \pgfmathsetmacro{\nx}{\cx+\w+\gap}\xdef\cx{\nx}%
  }%
  \node[chip, minimum width=\Tw pt] (#1t) at (\Tx,#2) {Tensors};
  \draw[flow] (#1p\nch.east) -- (#1t.west);
  \draw[flow] (#1t.east) -- (406,#2);
}
\figonepipe{nlp}{40.25}{Input text, Prompt format, Tokenization, Label masking}
\figonepipe{vis}{28.25}{Input image, Spatial augment, Normalization}
\figonepipe{med}{16.25}{Input volume, Orient/Space, Label format, Spatial crop, Intensity norm.}
\figonepipe{spe}{4.25}{Input audio, Resample, STFT, Padding/Mask}

\end{tikzpicture}

}
    \caption{Data pipelines in our experiments. Circled numbers mark the bottleneck each stage can cause (\autoref{fig:arch_diagram}).}
    \label{fig:data_pipelines}
\end{figure*}

Deep learning has expanded into numerous applications over the last decade,
each characterized by distinct model architectures, modalities, sample sizes
and file formats. We study four primary domains: Natural Language Processing (NLP), Computer Vision and Image Classification (CV), Medical Imaging and Audio and Speech Processing. 
\autoref{tab:domain-characteristics} summarizes how they differ in the data
they train on, how that data is stored and decoded, and the models they use,
and \autoref{fig:data_pipelines} shows the preprocessing steps of each
pipeline.

\begin{table}[h]
\centering
\caption{Characteristics of the four DL domains studied.}
\label{tab:domain-characteristics}
\footnotesize
\setlength{\tabcolsep}{3.5pt}
\renewcommand{\arraystretch}{1.1}
\newcolumntype{L}[1]{>{\raggedright\arraybackslash}p{#1}}
\begin{tabular}{@{}l L{1.50cm} L{1.75cm} L{1.35cm} L{1.9cm}@{}}
\toprule
\textbf{Domain} & \textbf{Data (sample)} & \textbf{Formats} & \textbf{Decoding} & \textbf{Typical models} \\
\midrule
NLP     & Text\newline{\color{black!60}B--KiB}            & Arrow, Parquet,\newline JSONL, CSV   & Text parse / row lookup   & Decoder-only\newline Transformers \\
Vision  & Images\newline{\color{black!60}$\sim$100\,KiB}  & JPEG, PNG,\newline WebDataset        & Image\newline decode          & CNNs, ViTs,\newline hybrid SSMs \\
Medical & 3D scans\newline{\color{black!60}1--100s\,MiB}  & NIfTI, DICOM\newline (often gzipped) & Decompress\newline + parse    & 3D U-Nets,\newline Swin-based \\
Speech  & Audio\newline{\color{black!60}$\sim$200--400\,KiB}   & WAV, FLAC,\newline MP3               & Audio codec\newline decoding  & Encoder--decoder\newline Transformers \\
\bottomrule
\end{tabular}
\end{table}

\boldfirst{Natural Language Processing (NLP):} Transformer-based language models are first pretrained on massive, unlabeled text datasets using self-supervised tasks like masked or autoregressive modeling. After text is processed via subword tokenizers (like BPE or WordPiece) and batched into fixed-length sequences, models undergo supervised fine-tuning (SFT) on instruction--response pairs (e.g., Alpaca~\cite{taori_stanford_2023}) and human preference alignment (RLHF or DPO).

\boldfirst{Computer Vision and Image Classification (CV):}  CV workloads (CNNs and ViTs) combine supervised pretraining on large labeled corpora (e.g., ImageNet) with self-supervised alternatives (contrastive learning, masked image modeling), stochastic augmentation, and channel-first tensor conversion.

\boldfirst{Medical Imaging:}  Medical imaging compounds this with volumetric data, since 3D U-Nets and CNN-Transformer hybrids train on 3D scans requiring orientation/spacing/intensity standardization and patch-wise sampling, with fewer labels and much larger samples than natural-image CV.

\boldfirst{Audio and Speech Processing:} Audio/speech workloads (Conformers, wav2vec 2.0/HuBERT-style self-supervised pretraining, CTC or attention-based decoding) operate on waveforms requiring sample-rate/channel standardization and feature extraction (e.g., mel-spectrograms) prior to sequence modeling.

\subsection{Data Pipelines in Training}
\label{sec:background:data_pipelines}
A DL training data pipeline is a producer-consumer system in which a dataloader continuously prepares batches while the accelerator consumes them for computation. To keep the accelerator busy, the two sides overlap. While one batch executes on the GPU, the next is retrieved from storage, decoded, and preprocessed on the CPU. Conceptually, every sample traverses the same stages regardless of domain, namely \textit{retrieval} from persistent storage, \textit{decoding} from its on-disk format, \textit{preprocessing} (augmentation, normalization, tokenization), \textit{transfer} to accelerator memory, and finally \textit{compute} as shown in \autoref{fig:data_pipelines}. These stages map directly onto a typical PyTorch implementation (\autoref{lst:code_example}, tags in the shades of \autoref{fig:data_pipelines}). The \texttt{Dataset}'s \texttt{\_\_getitem\_\_} method (1) performs retrieval and decoding for a single sample, while the \texttt{transform} pipeline (2) applies preprocessing. The \texttt{DataLoader} (3) orchestrates this across multiple worker processes, batching samples. The training loop (4) issues the host-to-device transfer (\texttt{.to(device)}) before invoking the forward, backward, and optimizer passes.
\begin{lstlisting}[caption={Example PyTorch Dataset and Training Loop{. Tags mark the stages \stagetag{stStorage}{S} storage, \stagetag[text=white]{stDecode}{D} decode, \stagetag{stPre}{P} preprocess, \stagetag[stXferPat]{stXfer}{T} transfer, \stagetag[text=white]{stCompute}{C} compute}}, label={lst:code_example}, float=tb, language=Python, basicstyle=\scriptsize\ttfamily, escapeinside={(*@}{@*)}]
import torch, io, ... # Import libs

# (1) Dataset implementation
class MyDataset(Dataset):
    def __init__(self, root, transform):
        self.samples = ImageFolder(root=root).samples
        self.transform = transform
    
    def __len__(self): # (required)
        return len(self.samples)
    
    def __getitem__(self, idx): # (required)
        file_path, label = self.samples[idx]
        with open(file_path, 'rb') as f:
            raw_bytes = f.read() # Buffered IO (*@\stagetag{stStorage}{S}@*)
        image = Image.open(io.BytesIO(raw_bytes)).convert("RGB") 
        image.load() # Decode using PIL (*@\stagetag[text=white]{stDecode}{D}@*)
        return self.transform(image), label (*@\stagetag{stPre}{P}@*)
        
# (2) Preprocessing transformations
transform = transforms.Compose([
    transforms.RandomResizedCrop(224), transforms.ToTensor()
])

# (3) Data loader
loader = DataLoader(
    MyDataset("path/to/data", transform), batch_size=64, 
    num_workers=4, pin_memory=True, persistent_workers=True,
    shuffle=True
)

# (4) Training loop
device = torch.device("cuda")
model = models.resnet50().to(device)
criterion = nn.CrossEntropyLoss()
optimizer = torch.optim.AdamW(model.parameters(), lr=1e-3)

for epoch in range(100):
    for images, labels in loader:
        images = images.to(device, non_blocking=True) (*@\stagetag[stXferPat]{stXfer}{T}@*)
        labels = labels.to(device, non_blocking=True)
        optimizer.zero_grad()
        loss = criterion(model(images), labels) (*@\stagetag[text=white]{stCompute}{C}@*)
        loss.backward(); optimizer.step()
\end{lstlisting}

\subsection{Software Stack and Data Path}
\label{sec:background:software_stack}
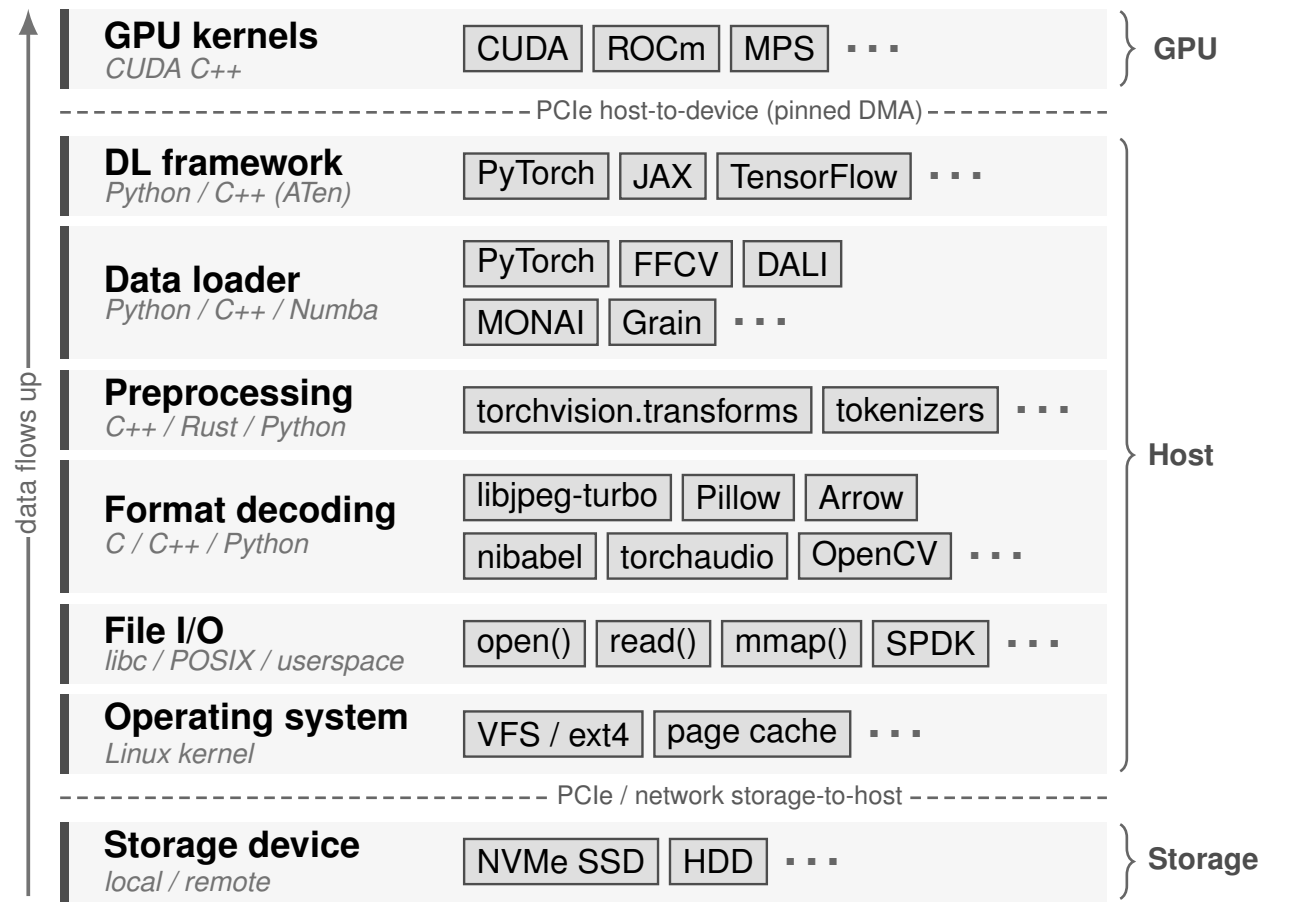
\begin{figure}
    \centering
    {\fontfamily{phv}\selectfont

\definecolor{cGPU}{HTML}{4D4D4D}
\definecolor{cFW}{HTML}{4D4D4D}
\definecolor{cPre}{HTML}{4D4D4D}
\definecolor{cDec}{HTML}{4D4D4D}
\definecolor{cIO}{HTML}{4D4D4D}
\definecolor{cOS}{HTML}{4D4D4D}
\definecolor{cSto}{HTML}{4D4D4D}
\definecolor{cLeg}{HTML}{4D4D4D}

\begin{tikzpicture}[
  font=\fontsize{6.2}{7}\selectfont,
  x=1pt,y=1pt,
  chip/.style={draw=#1, sharp corners, line width=0.45pt,
               inner xsep=2.4pt, inner ysep=1.3pt, minimum height=8.5pt,
               fill=#1!18, text=black, on chain, anchor=west},
  other/.style={draw=#1!50, sharp corners, line width=0.45pt, dash pattern=on 1.3pt off 0.9pt,
               inner xsep=2.4pt, inner ysep=1.3pt, minimum height=8.5pt,
               fill=white, text=black!60, on chain, anchor=west},
  lbl/.style={anchor=west, font=\fontsize{6.8}{7.5}\selectfont\bfseries},
  lang/.style={anchor=west, font=\fontsize{5.2}{6}\selectfont\itshape, text=black!55},
  lst/.style={circle, draw=black!70, line width=0.4pt, inner sep=0.3pt,
              minimum size=6.4pt, font=\fontsize{4.8}{5}\selectfont\bfseries},
  bn/.style={font=\fontsize{5.6}{6}\selectfont\bfseries, text=black!75},
  more/.style={on chain, anchor=west, inner xsep=1pt, inner ysep=1.2pt, minimum height=8.6pt, text=black!60, font=\bfseries},
  node distance=1.8pt,
]

\def\W{241}         
\def\Lx{12}          
\def\Cx{83}          
\def\Bx{215}        
\def\Dx{208.5}         

\newcommand{\stackrow}[6]{%
  \fill[black!3.5] (\Lx-3,#1) rectangle (\Bx-9,#1-#2);
  \fill[#3] (\Lx-3,#1) rectangle (\Lx-1.4,#1-#2);
  \node[lbl]  at (\Lx+#6,#1-#2/2+2.6) {#4};
  \node[lang] at (\Lx+#6,#1-#2/2-3.6) {#5};
}

\def\yA{210}\def\yB{186}\def\yC{169}\def\yD{142}\def\yE{125}\def\yF{98}\def\yG{81}\def\yH{57}

\stackrow{\yA}{15}{cGPU}{GPU kernels}{CUDA C++}{2}
\stackrow{\yB}{15}{cFW}{DL framework}{Python / C++ (ATen)}{2}
\stackrow{\yC}{25}{cFW}{Data loader}{Python / C++ / Numba}{2}
\stackrow{\yD}{15}{cPre}{Preprocessing}{C++ / Rust / Python}{2}
\stackrow{\yE}{25}{cDec}{Format decoding}{C / C++ / Python}{2}
\stackrow{\yF}{15}{cIO}{File I/O}{libc / POSIX / userspace}{2}
\stackrow{\yG}{15}{cOS}{Operating system}{Linux kernel}{2}
\stackrow{\yH}{15}{cSto}{Storage device}{local / remote}{2}

\begin{scope}[start chain=r1 going right]
  \node[on chain, inner sep=0pt, minimum size=0pt, outer sep=0pt, anchor=west] at (\Cx,\yA-7.5) {};
  \node[chip=cGPU]{CUDA}; \node[chip=cGPU]{ROCm}; \node[chip=cGPU]{MPS}; \node[more]{\dots};
\end{scope}
\begin{scope}[start chain=r2 going right]
  \node[on chain, inner sep=0pt, minimum size=0pt, outer sep=0pt, anchor=west] at (\Cx,\yB-7.5) {};
  \node[chip=cFW]{PyTorch}; \node[chip=cFW]{JAX}; \node[chip=cFW]{TensorFlow}; \node[more]{\dots};
\end{scope}
\begin{scope}[start chain=r3a going right]
  \node[on chain, inner sep=0pt, minimum size=0pt, outer sep=0pt, anchor=west] at (\Cx,\yC-7) {};
  \node[chip=cFW]{PyTorch}; \node[chip=cFW]{FFCV}; \node[chip=cFW]{DALI};
\end{scope}
\begin{scope}[start chain=r3b going right]
  \node[on chain, inner sep=0pt, minimum size=0pt, outer sep=0pt, anchor=west] at (\Cx,\yC-18) {};
  \node[chip=cFW]{MONAI}; \node[chip=cFW]{Grain}; \node[more]{\dots};
\end{scope}
\begin{scope}[start chain=r4 going right]
  \node[on chain, inner sep=0pt, minimum size=0pt, outer sep=0pt, anchor=west] at (\Cx,\yD-7.5) {};
  \node[chip=cPre]{torchvision.transforms}; \node[chip=cPre]{tokenizers}; \node[more]{\dots};
\end{scope}
\begin{scope}[start chain=r5a going right]
  \node[on chain, inner sep=0pt, minimum size=0pt, outer sep=0pt, anchor=west] at (\Cx,\yE-7) {};
  \node[chip=cDec]{libjpeg-turbo}; \node[chip=cDec]{Pillow}; \node[chip=cDec]{Arrow};
\end{scope}
\begin{scope}[start chain=r5b going right]
  \node[on chain, inner sep=0pt, minimum size=0pt, outer sep=0pt, anchor=west] at (\Cx,\yE-18) {};
  \node[chip=cDec]{nibabel}; \node[chip=cDec]{torchaudio}; \node[chip=cDec]{OpenCV}; \node[more]{\dots};
\end{scope}
\begin{scope}[start chain=r6 going right]
  \node[on chain, inner sep=0pt, minimum size=0pt, outer sep=0pt, anchor=west] at (\Cx,\yF-7.5) {};
  \node[chip=cIO]{open()}; \node[chip=cIO]{read()}; \node[chip=cIO]{mmap()}; \node[chip=cIO]{SPDK}; \node[more]{\dots};
\end{scope}
\begin{scope}[start chain=r7 going right]
  \node[on chain, inner sep=0pt, minimum size=0pt, outer sep=0pt, anchor=west] at (\Cx,\yG-7.5) {};
  \node[chip=cOS]{VFS / ext4}; \node[chip=cOS]{page cache}; \node[more]{\dots};
\end{scope}
\begin{scope}[start chain=r8 going right]
  \node[on chain, inner sep=0pt, minimum size=0pt, outer sep=0pt, anchor=west] at (\Cx,\yH-7.5) {};
  \node[chip=cSto]{NVMe SSD}; \node[chip=cSto]{HDD}; \node[more]{\dots};
\end{scope}

\draw[black!60, line width=0.5pt, dash pattern=on 2pt off 1.2pt]
  (\Lx-3,\yA-19.5) -- (\Bx-9,\yA-19.5);
\node[fill=white, inner sep=0.8pt, font=\fontsize{4.8}{5}\selectfont, text=black!65]
  at (\Cx+52,\yA-19.5) {PCIe host-to-device (pinned DMA)};
\draw[black!60, line width=0.5pt, dash pattern=on 2pt off 1.2pt]
  (\Lx-3,\yG-19.5) -- (\Bx-9,\yG-19.5);
\node[fill=white, inner sep=0.8pt, font=\fontsize{4.8}{5}\selectfont, text=black!65]
  at (\Cx+52,\yG-19.5) {PCIe / network storage-to-host};

\begin{scope}[decoration={brace,amplitude=2.5pt}, line width=0.5pt, black!60]
  \draw[decorate] (\Dx,\yA-0.5) -- (\Dx,\yA-14.5)
     node[midway, right=3pt, rotate=0, font=\fontsize{5.6}{6}\selectfont\bfseries, text=black!70]{GPU};
  \draw[decorate] (\Dx,\yB-0.5) -- (\Dx,\yG-14.5)
     node[midway, right=2pt, font=\fontsize{5.6}{6}\selectfont\bfseries, text=black!70]{Host};
  \draw[decorate] (\Dx,\yH-0.5) -- (\Dx,\yH-14.5)
     node[midway, right=2pt, font=\fontsize{5.6}{6}\selectfont\bfseries, text=black!70]{Storage};
\end{scope}

\draw[-latex, black!55, line width=0.7pt] (3,\yH-14) -- (3,\yA);
\node[rotate=90, font=\fontsize{5.2}{6}\selectfont, text=black!60, fill=white, inner sep=0.6pt]
  at (3,{(\yA+\yH-14)/2}) {data flows up};

\end{tikzpicture}

}
    \caption{Software stack traversed by a training sample from persistent storage to GPU memory.}
    \label{fig:software_stack}
\end{figure}

A file-backed training sample passes through the layers of \autoref{fig:software_stack} on its way from persistent storage to GPU memory, described here from the bottom up.

\boldfirst{Storage Device.} At the bottom, NVMe SSDs, HDDs, or remote storage execute the physical I/O, and data reaches host memory over PCIe or, for remote storage, the network.

\boldfirst{Operating System.} In the kernel, requests traverse the virtual file system, the filesystem (e.g., ext4 or XFS), the page cache, the block layer, and the device driver. The page cache serves data that is already in memory, and with \texttt{mmap()}, page faults populate memory on demand instead of explicit \texttt{read()} calls.

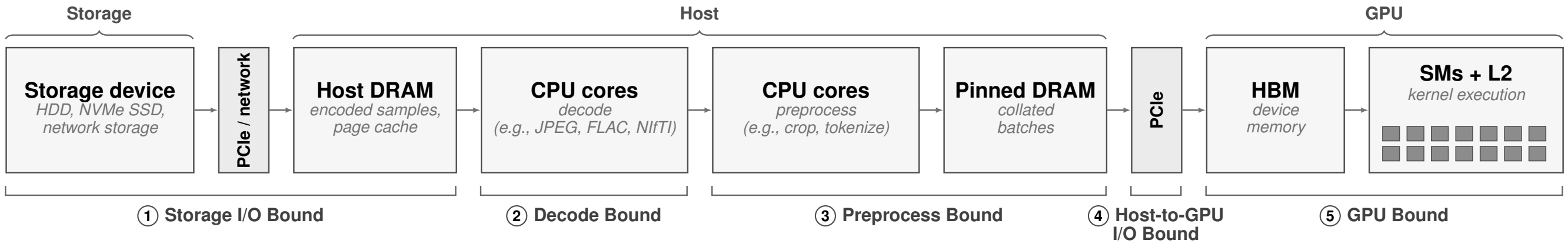
\begin{figure*}[t]
    \centering
    {\fontfamily{phv}\selectfont

\begin{tikzpicture}[x=1pt,y=1pt,font=\fontsize{6.2}{7}\selectfont,
  comp/.style={draw=black!70, line width=0.45pt, sharp corners, fill=black!3.5,
               align=center, inner sep=1.5pt, minimum height=40pt, anchor=south west},
  link/.style={draw=black!70, line width=0.45pt, sharp corners, fill=black!8,
               minimum width=16pt, minimum height=40pt, inner sep=0pt, anchor=south west},
  ttl/.style={font=\fontsize{6.8}{7.5}\selectfont\bfseries},
  sub/.style={font=\fontsize{5.2}{6}\selectfont\itshape, text=black!55},
  flow/.style={-{Latex[length=2.4pt]}, black!55, line width=0.5pt},
  grp/.style={text height=4.6pt, text depth=0.6pt, font=\fontsize{5.6}{6}\selectfont\bfseries, text=black!70},
  lst/.style={circle, draw=black!70, line width=0.4pt, inner sep=0.3pt,
              minimum size=6.4pt, font=\fontsize{4.8}{5}\selectfont\bfseries},
]

\newcommand{\figthreecomp}[6]{%
  \node[comp, minimum width=#3pt, fill=#6] (#1) at (#2,20)
    {\begin{tabular}{@{}c@{}}{\fontsize{6.8}{7.5}\selectfont\bfseries #4}\\[-0.5pt]
     {\renewcommand{\arraystretch}{0.8}\begin{tabular}{@{}>{\fontsize{5.2}{6}\selectfont\itshape\color{black!55}}c@{}}#5\end{tabular}}\end{tabular}};
}

\newcommand{\figthreelnk}[4]{%
  \node[link, fill=#4] (#1) at (#2,20) {};
  \node[rotate=90, font=\fontsize{5.4}{6}\selectfont\bfseries] at (#1.center) {#3};
}

\figthreecomp{sto}{4}{60}{Storage device}{HDD, NVMe SSD,\\network storage}{black!3.5}
\figthreelnk{l1}{72}{PCIe / network}{black!8}
\figthreecomp{dram}{96}{52}{Host DRAM}{encoded samples,\\page cache}{black!3.5}
\figthreecomp{dec}{156}{66}{CPU cores}{decode\\(e.g., JPEG, FLAC, NIfTI)}{black!3.5}
\figthreecomp{pre}{230}{66}{CPU cores}{preprocess\\(e.g., crop, tokenize)}{black!3.5}
\figthreecomp{pin}{304}{52}{Pinned DRAM}{collated\\batches}{black!3.5}
\figthreelnk{l2}{364}{PCIe}{black!8}
\figthreecomp{hbm}{388}{44}{HBM}{device\\memory}{black!3.5}

\node[comp, minimum width=62pt, fill=black!3.5] (sm) at (440,20) {};
\node[ttl, anchor=north] at ($(sm.north)+(0,-2.5)$) {SMs + L2};
\node[sub, anchor=north] at ($(sm.north)+(0,-10)$) {kernel execution};
\foreach \c in {0,...,6} \foreach \r in {0,1} {
  \draw[black!70, line width=0.3pt, fill=black!45]
    ($(sm.south west)+(4.5+\c*7.8,4+\r*6.4)$) rectangle +(5.4,4.6);
}

\draw[flow] (sto.east)  -- (l1.west);
\draw[flow] (l1.east)   -- (dram.west);
\draw[flow] (dram.east) -- (dec.west);
\draw[flow] (dec.east)  -- (pre.west);
\draw[flow] (pre.east)  -- (pin.west);
\draw[flow] (pin.east)  -- (l2.west);
\draw[flow] (l2.east)   -- (hbm.west);
\draw[flow] (hbm.east)  -- (sm.west);

\begin{scope}[decoration={brace,amplitude=2.5pt}, line width=0.45pt, black!60]
  \draw[decorate] (4,63)   -- (64,63)  node[midway, above=2.5pt, grp]{Storage};
  \draw[decorate] (96,63)  -- (356,63) node[midway, above=2.5pt, grp]{Host};
  \draw[decorate] (388,63) -- (502,63) node[midway, above=2.5pt, grp]{GPU};
\end{scope}

\newcommand{\figthreebn}[6]{%
  \draw[#6, line width=0.45pt] (#1,16) -- (#1,13) -- (#2,13) -- (#2,16);
  \node[inner sep=0pt, anchor=north] at ({(#1+#2)/2},10) {%
    \begin{tabular}{@{}c@{}}%
      \tikz[baseline=(c.base)]\node[lst](c){#3};\hspace{2pt}%
      {\fontsize{5.8}{6.5}\selectfont\bfseries\color{black!75}#4}\\[-1pt]
      {\fontsize{5.8}{6.5}\selectfont\bfseries\color{black!75}#5}%
    \end{tabular}};
}
\figthreebn{4}{148}{1}{Storage I/O Bound}{}{black!60}
\figthreebn{156}{222}{2}{Decode Bound}{}{black!60}
\figthreebn{230}{356}{3}{Preprocess Bound}{}{black!60}
\figthreebn{364}{380}{4}{Host-to-GPU}{I/O Bound}{black!60}
\figthreebn{388}{502}{5}{GPU Bound}{}{black!60}

\end{tikzpicture}

}
    \caption{Architecture overview and bottleneck classification.}
    \label{fig:arch_diagram}
\end{figure*}

\boldfirst{File I/O.} Userspace programs reach the operating system through POSIX calls such as \texttt{read()} or through memory-mapped files (\texttt{mmap()}).\footnote{Exceptions include userspace I/O libraries like SPDK~\cite{spdk}, which enable kernel bypass.}

\boldfirst{Format Decoding.} Libraries such as libjpeg-turbo, Pillow, Arrow, nibabel, or torchaudio parse, decompress, and materialize the on-disk format into arrays.

\boldfirst{Preprocessing.} After decoding, samples are transformed, for example by random cropping, normalization, or tokenization, usually on the CPU with libraries such as \texttt{torchvision.transforms} or tokenizers. Because augmentation is stochastic and repeated every epoch, it cannot be precomputed and is a common host-side bottleneck.

\boldfirst{Data Loader.} PyTorch's \texttt{Dataset} defines how samples are exposed (\texttt{\_\_len\_\_()}, \texttt{\_\_getitem\_\_()}). The \texttt{DataLoader} calls it, shuffles samples, collates them into batches, and, with multiple workers, runs the dataset in separate processes. FFCV, DALI, MONAI, and Grain replace or extend this layer. The \texttt{Dataset} implementation and the loader configuration together determine the I/O behavior.

\boldfirst{DL Framework.} PyTorch, JAX, or TensorFlow expose models in Python and dispatch operators through a native backend (ATen in PyTorch), which allocates tensors. With \texttt{pin\_memory=True}, batches are copied into page-locked host memory, from which the CUDA runtime transfers them to GPU memory by DMA over PCIe, asynchronously and without staging buffers.

\boldfirst{GPU Kernels.} At the top, the forward and backward passes run as GPU kernels (CUDA, ROCm, or MPS). They start only once a batch is in GPU memory.
The whole path is hidden behind the few lines of \autoref{lst:code_example}\footnote{We study this standard, CPU-centric path, plus DALI~\cite{dali}, which decodes and augments on the GPU. GPU Direct Storage~\cite{gds, gdseval} and GPU-initiated I/O~\cite{scada, torp_path_2025, qureshi_gpu-initiated_2023} bypass the CPU and are out of scope.}.

\subsection{Bottleneck Classification}
\label{sec:background:performance_hardware}

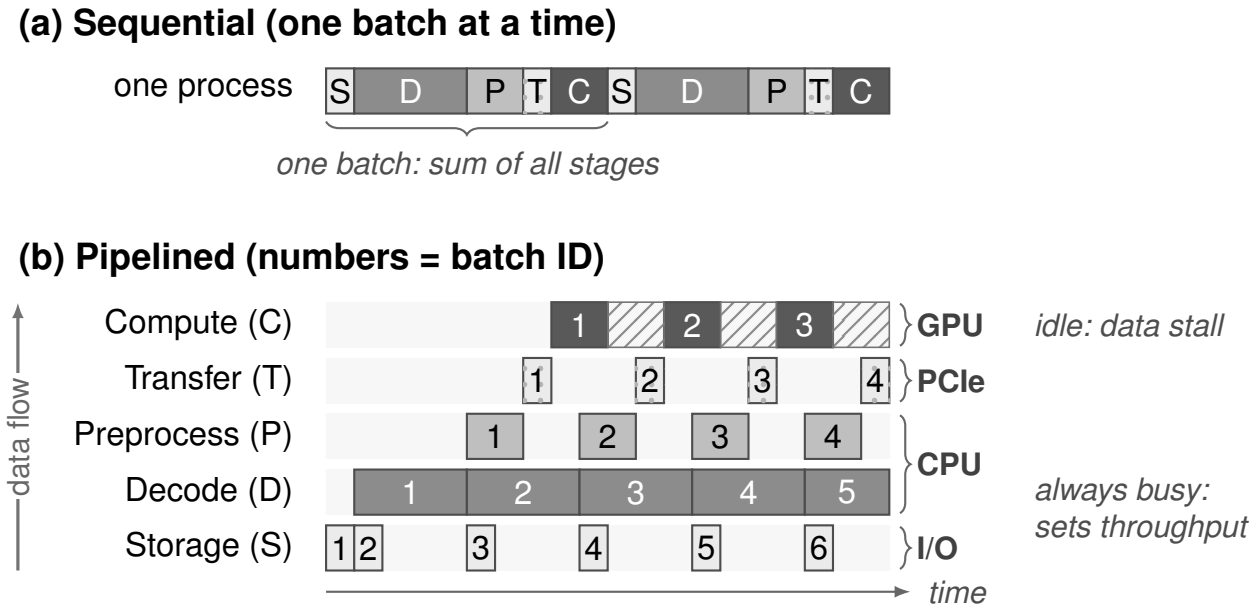
\begin{figure}[tb]
    \centering
    {\fontfamily{phv}\selectfont
\begin{tikzpicture}[x=1pt,y=1pt,font=\fontsize{6.2}{7}\selectfont,
  blk/.style={draw=black!70, line width=0.4pt, sharp corners},
  lane/.style={anchor=east, font=\fontsize{6.2}{7}\selectfont},
  ttl/.style={anchor=west, font=\fontsize{6.8}{7.5}\selectfont\bfseries},
  note/.style={font=\fontsize{5.8}{6.5}\selectfont\itshape, text=black!70}]
\node[ttl] at (0,122) {(a) Sequential (one batch at a time)};
\node[lane] at (58,109.75) {one process};
\draw[blk, fill=stStorage] (61.0,114) rectangle (66.3,105.5) node[midway, text=black, inner sep=0pt]{S};
\draw[blk, fill=stDecode] (66.3,114) rectangle (87.5,105.5) node[midway, text=white, inner sep=0pt]{D};
\draw[blk, fill=stPre] (87.5,114) rectangle (98.1,105.5) node[midway, text=black, inner sep=0pt]{P};
\draw[blk, fill=stXfer, stXferPat] (98.1,114) rectangle (103.4,105.5) node[midway, text=black, inner sep=0pt]{T};
\draw[blk, fill=stCompute] (103.4,114) rectangle (114.0,105.5) node[midway, text=white, inner sep=0pt]{C};
\draw[decorate,decoration={brace,amplitude=2.5pt,mirror},black!60,line width=0.4pt] (61.0,104.0) -- (114.0,104.0) node[midway,below=2.5pt,note]{one batch: sum of all stages};
\draw[blk, fill=stStorage] (114.0,114) rectangle (119.3,105.5) node[midway, text=black, inner sep=0pt]{S};
\draw[blk, fill=stDecode] (119.3,114) rectangle (140.5,105.5) node[midway, text=white, inner sep=0pt]{D};
\draw[blk, fill=stPre] (140.5,114) rectangle (151.1,105.5) node[midway, text=black, inner sep=0pt]{P};
\draw[blk, fill=stXfer, stXferPat] (151.1,114) rectangle (156.39999999999998,105.5) node[midway, text=black, inner sep=0pt]{T};
\draw[blk, fill=stCompute] (156.39999999999998,114) rectangle (167.0,105.5) node[midway, text=white, inner sep=0pt]{C};
\node[ttl] at (0,78) {(b) Pipelined (numbers = batch ID)};
\fill[black!3] (61,28.0) rectangle (167.0,19.5);
\node[lane] at (58,23.75) {Storage (S)};
\draw[blk, fill=stStorage] (61.0,28.0) rectangle (66.3,19.5) node[midway, text=black, inner sep=0pt]{1};
\draw[blk, fill=stStorage] (66.3,28.0) rectangle (71.6,19.5) node[midway, text=black, inner sep=0pt]{2};
\draw[blk, fill=stStorage] (87.5,28.0) rectangle (92.8,19.5) node[midway, text=black, inner sep=0pt]{3};
\draw[blk, fill=stStorage] (108.69999999999999,28.0) rectangle (114.0,19.5) node[midway, text=black, inner sep=0pt]{4};
\draw[blk, fill=stStorage] (129.89999999999998,28.0) rectangle (135.2,19.5) node[midway, text=black, inner sep=0pt]{5};
\draw[blk, fill=stStorage] (151.1,28.0) rectangle (156.39999999999998,19.5) node[midway, text=black, inner sep=0pt]{6};
\fill[black!3] (61,38.5) rectangle (167.0,30.0);
\node[lane] at (58,34.25) {Decode (D)};
\draw[blk, fill=stDecode] (66.3,38.5) rectangle (87.5,30.0) node[midway, text=white, inner sep=0pt]{1};
\draw[blk, fill=stDecode] (87.5,38.5) rectangle (108.69999999999999,30.0) node[midway, text=white, inner sep=0pt]{2};
\draw[blk, fill=stDecode] (108.69999999999999,38.5) rectangle (129.89999999999998,30.0) node[midway, text=white, inner sep=0pt]{3};
\draw[blk, fill=stDecode] (129.89999999999998,38.5) rectangle (151.1,30.0) node[midway, text=white, inner sep=0pt]{4};
\draw[blk, fill=stDecode] (151.1,38.5) rectangle (167.0,30.0) node[midway, text=white, inner sep=0pt]{5};
\fill[black!3] (61,49.0) rectangle (167.0,40.5);
\node[lane] at (58,44.75) {Preprocess (P)};
\draw[blk, fill=stPre] (87.5,49.0) rectangle (98.1,40.5) node[midway, text=black, inner sep=0pt]{1};
\draw[blk, fill=stPre] (108.69999999999999,49.0) rectangle (119.3,40.5) node[midway, text=black, inner sep=0pt]{2};
\draw[blk, fill=stPre] (129.89999999999998,49.0) rectangle (140.5,40.5) node[midway, text=black, inner sep=0pt]{3};
\draw[blk, fill=stPre] (151.1,49.0) rectangle (161.7,40.5) node[midway, text=black, inner sep=0pt]{4};
\fill[black!3] (61,59.5) rectangle (167.0,51.0);
\node[lane] at (58,55.25) {Transfer (T)};
\draw[blk, fill=stXfer, stXferPat] (98.1,59.5) rectangle (103.4,51.0) node[midway, text=black, inner sep=0pt]{1};
\draw[blk, fill=stXfer, stXferPat] (119.3,59.5) rectangle (124.6,51.0) node[midway, text=black, inner sep=0pt]{2};
\draw[blk, fill=stXfer, stXferPat] (140.5,59.5) rectangle (145.8,51.0) node[midway, text=black, inner sep=0pt]{3};
\draw[blk, fill=stXfer, stXferPat] (161.7,59.5) rectangle (167.0,51.0) node[midway, text=black, inner sep=0pt]{4};
\fill[black!3] (61,70.0) rectangle (167.0,61.5);
\node[lane] at (58,65.75) {Compute (C)};
\draw[blk, fill=stCompute] (103.4,70.0) rectangle (114.0,61.5) node[midway, text=white, inner sep=0pt]{1};
\draw[blk, fill=stCompute] (124.6,70.0) rectangle (135.2,61.5) node[midway, text=white, inner sep=0pt]{2};
\draw[blk, fill=stCompute] (145.8,70.0) rectangle (156.39999999999998,61.5) node[midway, text=white, inner sep=0pt]{3};
\draw[black!55, line width=0.3pt, pattern=north east lines, pattern color=black!45] (114.0,70.0) rectangle (124.6,61.5);
\draw[black!55, line width=0.3pt, pattern=north east lines, pattern color=black!45] (135.2,70.0) rectangle (145.8,61.5);
\draw[black!55, line width=0.3pt, pattern=north east lines, pattern color=black!45] (156.39999999999998,70.0) rectangle (167.0,61.5);
\draw[-{Latex[length=3pt]}, black!55, line width=0.6pt] (3,19.5) -- (3,70);
\node[rotate=90, fill=white, inner sep=0.6pt, font=\fontsize{5.6}{6}\selectfont, text=black!60] at (3,44.75) {data flow};
\draw[decorate,decoration={brace,amplitude=2pt},black!60,line width=0.4pt] (169.0,69.7) -- (169.0,61.8) node[midway,right=3pt,inner sep=0pt,font=\fontsize{5.8}{6.5}\selectfont\bfseries,text=black!75]{GPU};
\draw[decorate,decoration={brace,amplitude=2pt},black!60,line width=0.4pt] (169.0,59.2) -- (169.0,51.3) node[midway,right=3pt,inner sep=0pt,font=\fontsize{5.8}{6.5}\selectfont\bfseries,text=black!75]{PCIe};
\draw[decorate,decoration={brace,amplitude=2pt},black!60,line width=0.4pt] (169.0,48.7) -- (169.0,30.3) node[midway,right=3pt,inner sep=0pt,font=\fontsize{5.8}{6.5}\selectfont\bfseries,text=black!75]{CPU};
\draw[decorate,decoration={brace,amplitude=2pt},black!60,line width=0.4pt] (169.0,27.7) -- (169.0,19.8) node[midway,right=3pt,inner sep=0pt,font=\fontsize{5.8}{6.5}\selectfont\bfseries,text=black!75]{I/O};
\node[note, anchor=west] at (191.0,34.25) {always busy:};
\node[note, anchor=west] at (191.0,27.25) {sets throughput};
\node[note, anchor=west] at (191.0,65.75) {idle: data stall};
\draw[-{Latex[length=3pt]}, black!60, line width=0.4pt] (61,15.5) -- (171.0,15.5) node[right, note]{time};
\end{tikzpicture}}
    \caption{Sequential vs.\ pipelined data path. (a) Without overlap, a batch takes the sum of all stages. (b) With overlap, the slowest stage sets the throughput and the others wait.}
    \label{fig:pipelining}
\end{figure}

\autoref{sec:background:software_stack} presents the data path as a linear
sequence of layers, but frameworks do not run these layers one batch at a
time. \autoref{fig:pipelining} contrasts the two ways the stages can
execute. In the \emph{sequential} case (\autoref{fig:pipelining}a), a
single process takes one batch through storage, decoding, preprocessing,
transfer, and compute before it starts the next batch. The time per batch
is then the sum of all stages. In practice, frameworks \emph{pipeline}
these stages (\autoref{fig:pipelining}b). While the GPU computes on one
batch, dataloader workers are already reading, decoding, and preprocessing
the next batches. Asynchronous CUDA streams also overlap host-to-device
copies with kernel execution. In PyTorch, each worker performs storage,
decoding, and preprocessing for its own batch, and many workers run in
parallel, which is equivalent to the pipeline in
\autoref{fig:pipelining}b, where throughput is limited by whichever
resource saturates first.

Pipelining changes what limits performance. When stages overlap, their
times no longer add up. Instead, the pipeline can only move as fast as its
slowest stage. That stage is always busy and determines training
throughput, and every stage after it waits for input. If the slowest stage
comes before the GPU, the GPU is one of the waiting stages and sits idle.
This is called a \textit{data stall}. In a well-balanced system, the GPU is
the slowest stage, every upstream stage stays ahead of it, and the
accelerator is fully used. Adding dataloader workers speeds up CPU-bound
stages by processing more samples in parallel, until another
stage becomes the slowest.

Our evaluation uses both views. Sequential execution isolates how much
time each stage costs (\autoref{sec:latency}). Pipelined execution shows
which stage actually limits end-to-end throughput (\autoref{sec:scaling}).
To pin down where the slowest stage is, we split the path from storage to
the accelerator into five bottlenecks. Each one corresponds to a component that can limit
throughput, as shown in \autoref{fig:arch_diagram}.
On the host side, four bottlenecks can keep data from reaching the
accelerator fast enough:
\boldfirst{\bnum{1} Storage I/O Bound:} The storage medium (e.g., HDD, NVMe SSD, or
network storage), or the link between it and host memory, cannot deliver data as
fast as the host requests it.
\boldfirst{\bnum{2} Decode Bound:} The CPU cannot decode samples from their on-disk
format (e.g., JPEG, FLAC, NIfTI) fast enough.
\boldfirst{\bnum{3} Preprocess Bound:} The CPU cannot augment, transform, or collate
decoded samples fast enough.
\boldfirst{\bnum{4} Host-to-GPU I/O Bound:} The interconnect cannot move
prepared batches from host DRAM to GPU memory (HBM) fast enough.
On the accelerator side:
\boldfirst{\bnum{5} GPU Bound:} Data arrives faster than the GPU can process it,
so the GPU is the slowest stage and the input pipeline keeps up. This is
the desired state. A GPU-bound workload can be classified further as
\emph{memory bound}, where kernels wait on data moving from HBM to the
compute units, or \emph{compute bound}, where the GPU's peak flop/s at the
given precision sets the limit. That split concerns how the model runs on
the GPU rather than how data reaches it. Since this paper studies the data
pipeline that feeds the accelerator, we treat the GPU as a single stage.

\subsection{Checkpointing and the Write Path}
\label{sec:background:checkpointing}
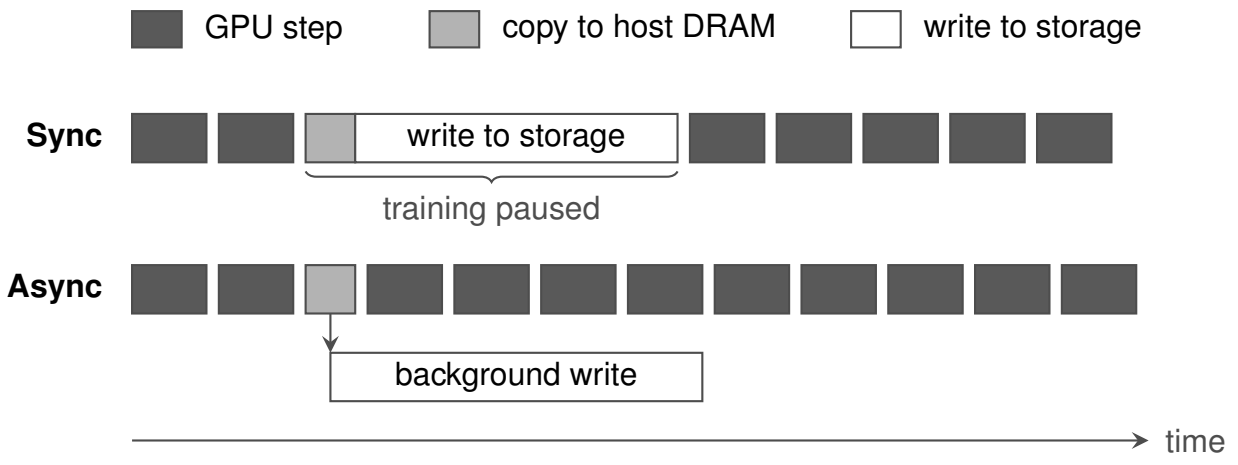
\begin{figure}[b]
    \centering
    {\fontfamily{phv}\selectfont

\begin{tikzpicture}[x=0.82mm,y=1mm,font=\scriptsize,
  stp/.style={draw=black!70,fill=black!65},
  cpy/.style={draw=black!70,fill=black!30},
  wrt/.style={draw=black!70,fill=white}]
  
  \draw[stp] (0,13) rectangle ++(4,2.4);  \node[anchor=west] at (4.5,14.2)  {GPU step};
  \draw[cpy] (24,13) rectangle ++(4,2.4); \node[anchor=west] at (28.5,14.2) {copy to host DRAM};
  \draw[wrt] (58,13) rectangle ++(4,2.4); \node[anchor=west] at (62.5,14.2) {write to storage};
  
  \node[anchor=east] at (-1,7) {\textbf{Sync}};
  \foreach \x in {0,7}{\draw[stp] (\x,5.4) rectangle ++(6,3.2);}
  \draw[cpy] (14,5.4) rectangle ++(4,3.2);
  \draw[wrt] (18,5.4) rectangle ++(26,3.2); \node at (31,7) {write to storage};
  \foreach \x in {45,52,59,66,73}{\draw[stp] (\x,5.4) rectangle ++(6,3.2);}
  \draw[black!70,decorate,decoration={brace,mirror,amplitude=2pt}] (14,4.8) -- (44,4.8)
       node[midway,below=1.5pt] {training paused};
  
  \node[anchor=east] at (-1,-3) {\textbf{Async}};
  \foreach \x in {0,7}{\draw[stp] (\x,-4.6) rectangle ++(6,3.2);}
  \draw[cpy] (14,-4.6) rectangle ++(4,3.2);
  \foreach \x in {19,26,33,40,47,54,61,68,75}{\draw[stp] (\x,-4.6) rectangle ++(6,3.2);}
  \draw[wrt] (16,-10.4) rectangle ++(30,3.2); \node at (31,-8.8) {background write};
  \draw[->,>=stealth,black!70] (16,-4.6) -- (16,-7.2);
  
  \draw[->,>=stealth,black!70] (0,-13) -- (82,-13) node[right] {time};
\end{tikzpicture}

}
    \caption{Synchronous vs.\ asynchronous checkpointing. Sync blocks training until the write completes, while async resumes after copying the state to host DRAM and writes in the background.}
    \label{fig:ckpt_timeline}
\end{figure}
Training periodically persists its state so that a failed job can resume.
A checkpoint holds the model weights, the optimizer state, and a scheduler
state that is negligible in size. With Adam~\cite{kingma_adam_2017}/AdamW~\cite{loshchilov_decoupled_2019},
the optimizer stores first and second moment estimates (momentum and
variance) for every parameter, so optimizer state makes up most of a
checkpoint, which is several times larger than the weights alone.
 
\boldfirst{Formats.} PyTorch's default format, Python Pickle~\cite{python_pickle}, serializes
arbitrary Python objects with no schema restrictions. This flexibility means
deserialization can execute arbitrary embedded code, a security risk with
untrusted checkpoints, and requires reading the entire file before any tensor
is usable. Safetensors from Hugging Face~\cite{hf_safetensors} instead restricts content to tensors
plus a lightweight JSON header (shapes, dtypes, byte offsets), which
eliminates code execution and enables zero-copy reads via memory-mapping, but
this stricter schema adds write-path overhead that Pickle avoids.
 
\boldfirst{Synchronous vs.\ asynchronous.} A synchronous checkpoint
(\texttt{torch\allowbreak.save()}) blocks training until the state has been
copied to host memory and persisted to storage
(\autoref{fig:ckpt_timeline}). An asynchronous checkpoint
with PyTorch Distributed Checkpoint (DCP)~\cite{pytorch_dcp} (\texttt{torch.\allowbreak distributed\allowbreak.checkpoint\allowbreak.async\allowbreak\_save()})
resumes training as soon as the state has been copied to host DRAM and
writes it in the background. This hides the write, but holds one host copy
per in-flight checkpoint.

\section{EXPERIMENTAL METHODOLOGY}
\label{sec:methodology}
\begin{figure*}[t]
    \centering
    \includegraphics[width=1.0\linewidth]{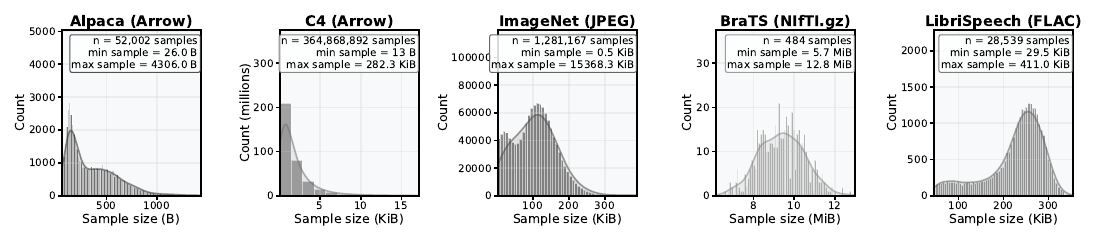}
    \caption{Per-sample size distribution for Alpaca, C4 (en), ImageNet, BraTS, and LibriSpeech.}
    \label{fig:dataset_distributions}
\end{figure*}

Our evaluation answers five research questions, each aimed at a decision a
practitioner makes when provisioning or configuring a training system:
\begin{rqbox}
\begin{description}[leftmargin=2.3em, labelwidth=2.0em, labelsep=0.3em, align=left, itemsep=1pt, parsep=0pt, topsep=0pt, partopsep=0pt, font=\normalfont\bfseries]
  \item[RQ1:] \textit{Where is time spent along the data path, and how does
    this differ across domains?} (\S\ref{sec:latency})
  \item[RQ2:] \textit{How much storage bandwidth and CPU parallelism does
    training need to keep the GPU busy?} (\S\ref{sec:scaling})
  \item[RQ3:] \textit{How do data formats and dataloaders affect throughput
    and host resources?} (\S\ref{sec:formats}--\ref{sec:memory})
  \item[RQ4:] \textit{Where does the pipeline stall outside steady-state
    training?} (\S\ref{sec:epoch-trans})
  \item[RQ5:] \textit{What does checkpointing cost, and how do format,
    frequency, and storage medium change it?} (\S\ref{sec:checkpointing})
\end{description}
\end{rqbox}
 
To isolate the bottlenecks described in \autoref{sec:background:performance_hardware} across a diverse set of real-world workloads, we use datasets and reference models from the MLCommons (MLPerf) training benchmarks~\cite{mattson_mlperf_2020} (e.g., ResNet-50 on ImageNet, C4 for language models). While MLPerf measures time-to-accuracy, we also vary file formats and break latency and throughput down by pipeline stage to locate the bottleneck of each domain and model.

\boldfirst{Experiment Hardware:} \textbf{Machine 1} features a single NVIDIA H100 PCIe GPU (80\,GiB HBM2e, 2~TB/s, PCIe Gen5 $\times$16), 2$\times$ Intel Xeon Silver 4410Y CPUs (48 threads in total, 2 NUMA nodes), 256\,GiB DRAM (16$\times$16\,GiB SK Hynix DDR5 RDIMM @ 4000~MT/s), and a local NVMe SSD (Samsung PM9A3 7.68~TB PCIe 4.0 $\times$4, 6.7/4.0~GB/s read/write~\cite{samsung_pm9a3_2022}). \textbf{Machine 2} is used for the checkpointing experiments on NFS (\autoref{sec:ckpt-freq}), has 2$\times$ NVIDIA H100 PCIe GPUs (80\,GiB, PCIe Gen5 $\times$16, 350W), 2$\times$ AMD EPYC 9354 CPUs (2 NUMA nodes), 768\,GiB DDR5 RAM, and remote storage via NFSv4.2 over RDMA (100Gb/s EDR InfiniBand) backed by an 11$\times$7.68~TB Micron 7450 NVMe RAID 5 array. \textbf{Machine 3} is used for the multi-GPU (DDP) experiment (\autoref{sec:scaling}), has 8$\times$ NVIDIA H100 SXM GPUs (80\,GiB HBM3, 3.35~TB/s, 700W TDP), 2$\times$ Intel Xeon Platinum 8480C CPUs (224 threads in total), 2\,TiB DDR5 host memory, and WekaFS storage, and runs NVIDIA DGX OS.
All other experiments run on Machine 1. 

\boldfirst{Software Frameworks:} We evaluate four dataloading frameworks that represent distinct approaches to the \textit{data stall} problem. PyTorch~\cite{paszke_pytorch_2019} (Python/C++) and MONAI~\cite{cardoso_monai_2022} (Python/C++/CUDA, for medical imaging) parallelize loading with worker processes. FFCV~\cite{leclerc_ffcv_2023} (Numba, for vision) runs its pipeline in threads to reduce Python-side overhead. NVIDIA DALI~\cite{dali} (C++/CUDA, for vision) uses threads and offloads decoding and augmentation to the GPU, as do parts of MONAI. NLP and Speech use the Hugging Face \texttt{datasets} library with the PyTorch \texttt{DataLoader}.

\boldfirst{Datasets and Formats:}
Our datasets range from small-sample NLP text to large-sample Medical Imaging volumes (\Cref{tab:format_comparison}, \autoref{fig:dataset_distributions}). 
In the NLP domain, we compare the storage efficiency and loading latency of \texttt{.arrow}, \texttt{.parquet}, \texttt{.jsonl}, and \texttt{.csv} formats using the Alpaca~\cite{taori_stanford_2023, wang_self-instruct_2023} and C4~\cite{raffel_exploring_2023} datasets. 
The Computer Vision benchmarks utilize ImageNet-1K to contrast standard file-per-sample formats (JPEG/PNG) against aggregate formats like WebDataset (\texttt{.tar}) and the specialized \texttt{.beton} format used by FFCV. 
Finally, Medical Imaging (BraTS) provides a high sample-size case, where a single sample (image and label volumes) averages 145\,MiB uncompressed and 9.4\,MiB gzipped.

\begin{table}[t]
    \centering
    \caption{Experimental setup for Text (Alpaca~\cite{taori_stanford_2023, wang_self-instruct_2023}, C4~\cite{raffel_exploring_2023}), Image (ImageNet-1K~\cite{deng_imagenet_2009}), Medical Image (BraTS~\cite{verdier_2024_2024}), and Audio (LibriSpeech~\cite{panayotov_librispeech_2015}) data formats.}
    \label{tab:format_comparison}
    \footnotesize
    \setlength{\tabcolsep}{2.5pt}
    \begin{tabularx}{\linewidth}{@{}>{\raggedright\arraybackslash}X l r r l l@{}}
        \toprule
        \textbf{Dataset} & \textbf{Format} & \textbf{Per sample} & \textbf{Total} & \textbf{Read by} & \textbf{Loader}\\
        \midrule
        \multirow{4}{\linewidth}{\raggedright \textit{Alpaca}\\ ($n$=52,002)}
        & Arrow  & 228.00 B  & 11.3 MiB  & HF Datasets & PyTorch \\
        & Parquet & 222.00 B & 11.0 MiB & HF Datasets & PyTorch \\
        & JSONL & 397.41 B & 19.7 MiB & HF Datasets & PyTorch \\
        & CSV & 358.79 B & 17.8 MiB & HF Datasets & PyTorch \\
        \textit{C4} (364.9\,M)
        & Arrow & 1,353 B & 460.33 GiB & HF Datasets & PyTorch \\
        \midrule
        \multirow{4}{\linewidth}{\raggedright \textit{ImageNet-1K}\\ ($n$=1,281,167)}
        & JPEG & 111.98 KiB  & 136.8 GiB & Pillow, DALI & PyTorch, DALI \\
        & PNG & 345.47 KiB & 422.1 GiB & Pillow & PyTorch \\
        & WebDataset & 115.73 KiB & 141.4 GiB & WebDataset & PyTorch \\
        & Beton\footnotemark & 588.86 KiB & 719.5 GiB & FFCV & FFCV \\
        \midrule
        \multirow{2}{\linewidth}{\raggedright \textit{BraTS}\\ ($n$=484)}
        & NIfTI & 144.75 MiB  & 68.4 GiB & MONAI & MONAI \\
        & NIfTI-GZ & 9.40 MiB  &  4.4 GiB & MONAI & MONAI \\
        \midrule
        \multirow{4}{\linewidth}{\raggedright \textit{LibriSpeech}\\ ($n$=28,539)}
        & Arrow  & 396.74 KiB &  10.8 GiB & HF Datasets & PyTorch \\
        & WAV & 396.77 KiB & 10.9 GiB & torchaudio & PyTorch \\
        & FLAC & 226.44 KiB &  6.2 GiB & torchaudio & PyTorch \\
        & MP3 & 250.52 KiB & 6.9 GiB & torchaudio & PyTorch \\
        \bottomrule
    \end{tabularx}
\end{table}
\footnotetext{The \texttt{.beton} file stores uncompressed images as 3-channel \texttt{uint8} tensors (one byte per channel per pixel), fully decoding each JPEG onto disk. FFCV provides a compression parameter that allows practitioners to keep a subset of the dataset as JPEGs while storing the rest uncompressed. We use the default configuration, which stores all images fully decoded.}

\boldfirst{Model Architectures:} The selection of models,
which Table \ref{tab:models_arch_size} details, spans several architecture families, including pure Transformers (Llama, ViT, Whisper), standard CNNs (ResNet, 3D-UNet), and emerging hybrid state space models (MambaVision). This variety allows us to vary the \textit{arithmetic intensity} of the execution phase on the GPU, which in turn affects  the throughput requirement in the upstream pipeline. We update all parameters of every model. The NLP and Speech models are fine-tuned from pretrained weights, while the Vision and Medical models are trained from random initialization.

\begin{table}[t]
    \centering
    \caption{Model Architecture, Overview, Data type, and Sizes (MiB). Dtype is the weight precision, and the fp32 models train with automatic mixed precision.}
    \label{tab:models_arch_size}
    \footnotesize
    \setlength{\tabcolsep}{2.4pt} 
    \begin{tabularx}{\columnwidth}{@{}ll >{\raggedright\arraybackslash}X rcrr@{}}
        \toprule
        \textbf{Domain} & \textbf{Model} & \textbf{Architecture} & \textbf{Params} & \textbf{Dtype} & \textbf{Size} & \textbf{Batch}\\ 
        \midrule
        \multirow{4}{*}{NLP} 
            & Llama-3.2-1B~\cite{grattafiori2024llama3herdmodels} & Transf. (Dec) & 1235.8M & bf16 & 2357.1 & 32\\
            & Llama-3.2-3B~\cite{grattafiori2024llama3herdmodels}\footnotemark & Transf. (Dec) & 3212.7M & bf16 & 6127.8 & 16\\
            & Qwen3-0.6B~\cite{yang_qwen3_2025}    & Transf. (Dec) & 596.0M  & bf16 & 1136.9 & 32 \\
            & Qwen3-1.7B~\cite{yang_qwen3_2025}    & Transf. (Dec) & 1720.6M & bf16 & 3281.7 & 16\\ 
        \midrule
        \multirow{3}{*}{\shortstack[l]{Image\\Class.}} 
            & ResNet-50~\cite{he_deep_2015}             & CNN                 & 25.6M   & fp32 & 97.5 & 1024  \\
            & ViT~\cite{dosovitskiy_image_2021}         & Transf. (Enc)   & 86.6M   & fp32 & 330.2 & 1024 \\
            & MambaVision~\cite{hatamizadeh_mambavision_2025} & Hybrid SSM & 31.8M   & fp32 & 121.3 & 1024 \\ 
        \midrule
        \multirow{3}{*}{\shortstack[l]{Medical\\Imaging}} 
            & 3D-UNet~\cite{cicek_3d_2016}   & CNN (U-Net)        & 4.8M    & fp32 & 18.3 & 32\\
            & SwinUNETR~\cite{hatamizadeh_swin_2022} & Transf. (Swin) & 62.2M   & fp32 & 237.2 & 1\\
            & SegResNet~\cite{myronenko_3d_2018}  & CNN & 4.7M    & fp32 & 17.9 & 4\\ 
        \midrule
        Speech 
            & Whisper-small~\cite{radford_robust_2022} & Transf. (Enc-Dec) & 241.7M  & fp32 & 922.1 & 64 \\ 
        \bottomrule
    \end{tabularx}

\end{table}
\footnotetext{Llama-3.2-3B is the largest of our models that fits on 80\,GiB HBM at batch size 16 (72.5\,GiB including parameters, gradients, optimizer state, and activations).}

\boldfirst{Metrics and Measurements:} To characterize the system performance across the entire I/O stack requires a multi-layered approach.

Storage layer activity is captured using \texttt{iostat}, through which we monitor \textit{read/write throughput} (MiB/s) and \textit{operation intensity} (IOPS). 
We also check
\textit{SSD write amplification}, 
which is the ratio of bytes written to flash to bytes written by the host, read from the drive's NVMe and OCP SMART logs before and after a run. \texttt{vmstat} and \texttt{psutil}~\cite{psutil} record CPU utilization and host DRAM usage. 

For GPU activity, we track performance using the NVIDIA Data Center GPU Manager (DCGM) and \texttt{nvidia-smi}. The metrics are \textit{SMACT} (percentage of SMs active), \textit{SMOCC} (average achieved occupancy of active warps on the SMs), \textit{TENSO} (percentage of cycles the tensor cores are active), and \textit{DRAMA} (percentage of time the device memory interface is active).

Beyond system-level telemetry, the dataloader is instrumented with timers and counters to measure application-specific performance. We specifically track \textit{sample throughput} (samples/s) to evaluate end-to-end efficiency.

We report data and memory sizes in binary units (KiB, MiB, GiB), while hardware bandwidths and SSD capacities are the vendor-specified decimal values. We define a \textit{batch} as a collection of data samples, a \textit{step} as the processing of a batch on the GPU, and an \textit{epoch} as the processing of the entire dataset exactly once. All experiments use random sampling unless stated otherwise. To isolate the effects of the operating system's page cache, we clear the cache before each trial on Machine 1 to establish a cold-start state.

\section{DATA LOADING CHARACTERIZATION}
\label{sec:data_loading}

\subsection{Pipeline Latency Differs Across Domains}
\label{sec:latency}
To characterize the underlying cost of each stage in the data path, we first perform a latency decomposition of a training batch, averaged over one epoch. Because dataloaders use multiple workers to mask I/O and preprocessing costs, the stages overlap, so we isolate them by executing the training loop in a synchronous, single-process mode (\texttt{num\_workers=0}, i.e., the main process loads the data itself). This configuration does not reach peak throughput, but it measures the time spent on disk retrieval, format decoding, augmentation, host-to-device transfer, and compute for each batch to reveal the characteristics of each domain.
\begin{figure}[t]
    \centering
    \includegraphics[width=1.0\linewidth]{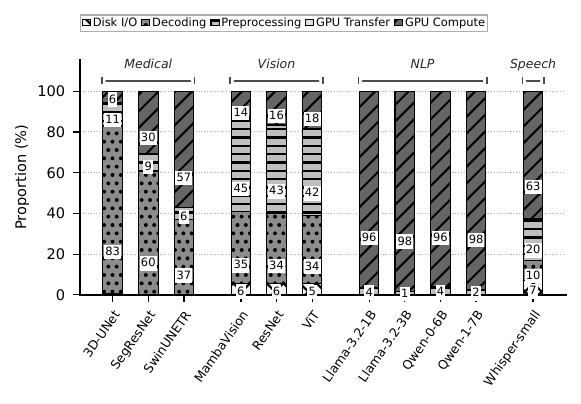}
    \caption{Time breakdown of different training stages across domains. Decoding and preprocessing dominate Medical and Vision (except SwinUNETR, 57\% GPU compute), while NLP is $\geq$96\% GPU compute.}
    \label{fig:latency_normalized}
\end{figure}
\begin{figure*}
    \centering
    \includegraphics[width=0.9\linewidth]{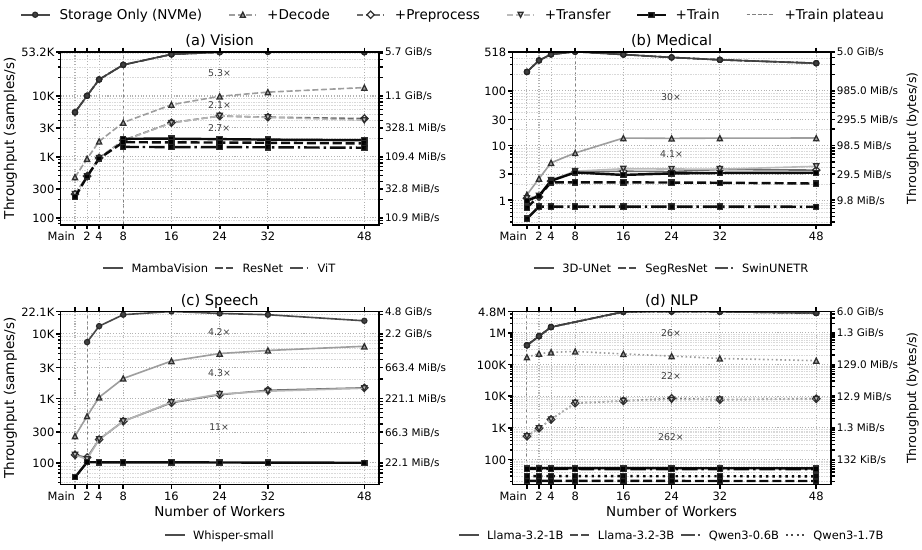}
    \caption{Throughput vs.\ worker count for each cumulative stage (\autoref{tab:pipeline_stages}). Storage Only through +Transfer shows one model per domain (ResNet-50, 3D-UNet, Whisper-small, Qwen3-1.7B). +Train shows all models. Labels between curves give the slowdown from one stage to the next at 24 workers; the dashed line marks where +Train reaches 95\% of its peak. \emph{Main}: \texttt{num\_workers=0}.}
    \label{fig:tput_stages_main_fig}
\end{figure*}

As shown in \autoref{fig:latency_normalized}, the distribution of execution time across batch-processing stages varies substantially between domains. In the Medical Imaging and Computer Vision domains, the latency is heavily skewed toward data decoding and preprocessing (Bottlenecks 2 and 3, \autoref{sec:background:performance_hardware}). For Medical Imaging specifically, the overhead of decompressing (Bottleneck 2) complex volumetric formats (e.g., \texttt{.nii.gz}) dominates the pipeline for 3D-UNet and SegResNet (83\% and 60\%), while SwinUNETR spends 57\% in GPU compute. In Vision, preprocessing (42--45\%) takes longer than decoding (34--35\%). On the other hand, NLP workloads show a very compute-dominant profile. Despite the overhead of tokenization, the arithmetic intensity of the Transformer models used in this domain is so high that the majority of the time is spent on compute (Bottleneck 5) even under suboptimal conditions without multi-worker data loading. Finally, the speech domain is mostly compute, with little time in preprocessing. 

\begin{insight}\label{ins:nlp-gpu}
NLP is GPU bound even without parallel data loading. All four NLP models spend 96--98\% of per-batch time in GPU compute using only the main process for data loading (zero dataloader workers).
\end{insight}
 
\begin{insight}\label{ins:decode}
Decoding and preprocessing dominate Vision and Medical. Decoding alone takes 83\% of per-batch time for 3D-UNet, 60\% for SegResNet, and 34--35\% for the Vision models. SwinUNETR is the exception, with GPU compute taking 57\% and decoding 37\%.
\end{insight}

\subsection{Storage Outpaces Training Demand}
\label{sec:scaling}
To prevent the pipeline from stalling, DL frameworks typically deploy \textit{multiple} dataloader workers, where each worker concurrently executes both disk I/O and preprocessing for different samples. We now evaluate how each pipeline scales with increasing worker counts. To measure the impact of each incremental step, we decompose the pipeline into five cumulative stages (detailed in \autoref{tab:pipeline_stages}).

\begin{table}[t]
\centering
\caption{Stages of the data ingestion and training pipeline.}
\label{tab:pipeline_stages}
\footnotesize
\setlength{\tabcolsep}{3.5pt}
\begin{tabular}{@{}lll@{}}
\toprule
\textbf{\#} &
\textbf{Pipeline Stage} & \textbf{Included Components} \\ \midrule
\textbf{1} & \textbf{Storage Only}   & SSD data retrieval + PCIe transfer. \\
\textbf{2} & \textbf{+ Decode}       & Adds format decoding. \\
\textbf{3} & \textbf{+ Preprocess}    & Adds CPU augmentations \& tensor conversions. \\
\textbf{4} & \textbf{+ Transfer}     & Adds copying each batch to GPU memory (HBM). \\
\textbf{5} & \textbf{+ Train}        & Adds GPU Computation (full end-to-end). \\ \bottomrule
\end{tabular}
\end{table}
\begin{table*}[t]
\centering
\footnotesize
\setlength{\tabcolsep}{0.5pt}
\linespread{0.95}\selectfont
\caption{Throughput and resource metrics by storage format at 16 workers for Vision (ResNet-50), Medical (3D-UNet), Speech (Whisper-small), and NLP (Llama-3.2-1B). CPU Mem is system-wide host memory in use, including shared memory, and the first and last 60 s are trimmed. Vision runs use the channels-last memory format to match FFCV.}
\label{tab:combined-metrics}%
\begin{tabular*}{\linewidth}{@{\extracolsep{\fill}}p{0.3cm}p{1.55cm}rcccccc@{}}
\toprule
 & \textbf{Format} & \textbf{Throughput $\uparrow$} & \textbf{CPU (\%) $\downarrow$} & \multicolumn{1}{c|}{\textbf{CPU Mem (GiB) $\downarrow$}} & \textbf{SMACT (\%) $\uparrow$} & \textbf{SMOCC (\%) $\uparrow$} & \textbf{DRAMA (\%) $\uparrow$} & \textbf{GPU Mem (GiB) $\downarrow$} \\
 & & \textbf{(Samp/s)} & \multicolumn{6}{c}{\textbf{Mean (Min--Max)}} \\
\cmidrule(r){1-2} \cmidrule(r){3-3} \cmidrule(l){4-5} \cmidrule(l){6-9}
\multirow{5}{*}{\rotatebox[origin=c]{90}{Vision}} & \mbox{JPEG} & 2,876.14 & 34.11 {\footnotesize (31.90--36.10)} & 68.27 {\footnotesize (65.15--71.73)} & 90.47 {\footnotesize (76.70--93.20)} & 45.13 {\footnotesize (38.10--46.40)} & 70.02 {\footnotesize (60.70--77.60)} & 53.71 {\footnotesize (53.71--53.71)} \\
 & \mbox{PNG} & 1,546.79 & 37.66 {\footnotesize (34.20--41.80)} & 62.25 {\footnotesize (59.90--66.96)} & 44.50 {\footnotesize (0.00--92.70)} & 22.37 {\footnotesize (0.00--46.80)} & 36.31 {\footnotesize (0.00--78.20)} & 53.71 {\footnotesize (53.71--53.71)} \\
 & \mbox{WebDataset} & 2,850.67 & 38.09 {\footnotesize (35.00--41.10)} & 66.98 {\footnotesize (62.90--70.14)} & 90.12 {\footnotesize (82.30--93.10)} & 44.97 {\footnotesize (41.30--46.20)} & 70.84 {\footnotesize (65.30--79.20)} & 53.89 {\footnotesize (53.89--53.89)} \\
 & \mbox{Beton (FFCV)} & \textbf{3,039.02} & 21.50 {\footnotesize (16.80--28.50)} & 24.44 {\footnotesize (23.17--25.52)} & \textbf{95.54 {\footnotesize (95.00--95.80)}} & \textbf{47.42 {\footnotesize (46.40--48.40)}} & \textbf{73.15 {\footnotesize (64.50--82.20)}} & \textbf{52.15 {\footnotesize (52.15--52.15)}} \\
 & \mbox{JPEG (DALI)} & 2,818.67 & \textbf{17.27 {\footnotesize (12.40--27.80)}} & \textbf{13.59 {\footnotesize (13.11--14.64)}} & 84.81 {\footnotesize (80.50--90.10)} & 42.66 {\footnotesize (40.80--44.70)} & 66.28 {\footnotesize (61.40--73.50)} & 57.59 {\footnotesize (57.24--57.75)} \\
\midrule
\multirow{2}{*}{\rotatebox[origin=c]{90}{Med.}} & \mbox{NIfTI} & \textbf{4.74} & \textbf{20.02 {\footnotesize (4.20--27.20)}} & \textbf{84.93 {\footnotesize (58.07--118.40)}} & \textbf{8.03 {\footnotesize (0.00--77.40)}} & \textbf{3.57 {\footnotesize (0.00--35.10)}} & \textbf{3.00 {\footnotesize (0.00--28.60)}} & 8.67 {\footnotesize (0.47--44.61)} \\
 & \mbox{NIfTI-GZ} & 4.21 & 24.23 {\footnotesize (4.20--28.60)} & 103.31 {\footnotesize (66.50--149.95)} & 4.91 {\footnotesize (0.00--93.00)} & 2.18 {\footnotesize (0.00--41.80)} & 1.82 {\footnotesize (0.00--31.20)} & \textbf{4.51 {\footnotesize (0.47--44.61)}} \\
\midrule
\multirow{3}{*}{\rotatebox[origin=c]{90}{Speech}} & \mbox{WAV} & 97.45 & 8.22 {\footnotesize (7.80--9.30)} & 23.09 {\footnotesize (21.64--23.82)} & 93.78 {\footnotesize (92.10--95.00)} & 46.62 {\footnotesize (45.10--48.10)} & 44.79 {\footnotesize (40.10--50.50)} & \textbf{63.82 {\footnotesize (60.17--67.20)}} \\
 & \mbox{FLAC} & 97.47 & \textbf{7.29 {\footnotesize (6.90--8.20)}} & \textbf{21.03 {\footnotesize (20.62--21.73)}} & 93.72 {\footnotesize (91.90--94.80)} & 46.58 {\footnotesize (44.70--48.20)} & 44.78 {\footnotesize (41.80--50.40)} & 63.83 {\footnotesize (60.17--67.20)} \\
 & \mbox{MP3} & \textbf{97.99} & 7.36 {\footnotesize (6.90--8.10)} & 22.79 {\footnotesize (21.83--23.43)} & \textbf{93.89 {\footnotesize (92.80--94.90)}} & \textbf{46.71 {\footnotesize (45.10--48.20)}} & \textbf{45.02 {\footnotesize (39.80--50.30)}} & 63.83 {\footnotesize (60.17--67.20)} \\
\midrule
\multirow{4}{*}{\rotatebox[origin=c]{90}{NLP}} & \mbox{Arrow} & 52.43 & 4.41 {\footnotesize (4.30--4.60)} & 7.32 {\footnotesize (7.31--7.37)} & 95.85 {\footnotesize (95.20--96.60)} & 43.29 {\footnotesize (41.00--45.50)} & 48.98 {\footnotesize (43.20--56.20)} & \textbf{74.97 {\footnotesize (74.97--74.97)}} \\
 & \mbox{CSV} & \textbf{52.66} & 4.45 {\footnotesize (4.30--5.40)} & 7.30 {\footnotesize (7.29--7.34)} & \textbf{95.85 {\footnotesize (95.20--96.50)}} & 43.33 {\footnotesize (41.00--45.70)} & \textbf{49.21 {\footnotesize (42.80--55.30)}} & \textbf{74.97 {\footnotesize (74.97--74.97)}} \\
 & \mbox{JSONL} & 52.60 & \textbf{4.41 {\footnotesize (4.30--5.20)}} & 7.31 {\footnotesize (7.29--7.38)} & 95.83 {\footnotesize (95.10--96.50)} & \textbf{43.34 {\footnotesize (40.00--45.60)}} & 49.18 {\footnotesize (42.90--56.40)} & \textbf{74.97 {\footnotesize (74.97--74.97)}} \\
 & \mbox{Parquet} & 52.50 & 4.43 {\footnotesize (4.30--5.00)} & \textbf{7.29 {\footnotesize (7.27--7.33)}} & 95.84 {\footnotesize (95.10--96.60)} & 43.32 {\footnotesize (40.60--45.60)} & 49.05 {\footnotesize (42.80--54.90)} & \textbf{74.97 {\footnotesize (74.97--74.97)}} \\
\bottomrule
\end{tabular*}
\end{table*}

 \autoref{fig:tput_stages_main_fig} shows that isolated storage throughput substantially overstates the rate sustained by the full input pipeline. At 24 workers, storage-only throughput is $30.4\times$, $131.6\times$, and $202.5\times$ higher than end-to-end training throughput for Vision (ResNet-50), Medical (3D-UNet), and Speech (Whisper-small), respectively. For NLP, the gap is five orders of magnitude.\footnote{For NLP, the storage benchmark reads the 1,024 \texttt{.arrow} files of C4 (about 460\,MiB each) and therefore measures file-level rather than sample-level throughput. Packed samples make these reads predominantly sequential, so the resulting storage-to-training ratio is not directly sample-comparable. The +Decode stage captures the cost of recovering individual samples from files.} These large relative drops do not indicate a CPU bottleneck for NLP. Even the slowest NLP stage, tokenization (about 8,500 samples/s for Qwen3-1.7B at 24 workers), is about 290$\times$ faster than training.
Introducing decoding reduces throughput by 76--97\% across the evaluated models, the largest single-stage drop in the Vision and Medical pipelines. Preprocessing causes an additional 52--95\% reduction (each model relative to its own decode stage, at 24 workers). GPU transfer has a limited effect.  Finally, adding training reduces throughput by 17\% for Medical, 63.4\% for Vision, 91.3\% for Speech, and 99.6\% for NLP.

To evaluate the scalability of our findings, we conduct an experiment on Machine 3, where we train (full pipeline) Llama-3.2-3B on the C4 dataset using DDP~\cite{li_pytorch_2020}, and training needs only 48.61 KiB/s of training data (36.79 samples/s) to keep 1 GPU saturated, 91.73 KiB/s (69.42 samples/s) for 2 GPUs, 181.01 KiB/s (136.99 samples/s) for 4 GPUs, and 357.78 KiB/s (270.78 samples/s) for 8 GPUs\footnote{Scaling efficiency is 94\%, 93\%, and 92\% on 2, 4, and 8 GPUs. 
This is due to
the gradient all-reduce between steps, during which each GPU exchanges 29--42\,GB/s over NVLink, and its tensor-core activity drops by 5--7\% relative to one GPU.}. Across 1 to 8 GPUs, SMACT stays at 91--94\% and each GPU needs about 45\,KiB/s (27.5\,KiB/s on Machine~1's H100 PCIe), so even 8 GPUs need four orders of magnitude less than the read throughput of one NVMe SSD (6.7\,GB/s). 

\begin{insight}\label{ins:storage}
Storage was not the bottleneck in any pipeline we evaluated. Storage I/O is at most 7\% of per-batch latency in every pipeline (\autoref{fig:latency_normalized}), and at 24 workers storage-only throughput exceeds end-to-end training throughput by 30$\times$ (Vision), 132$\times$ (Medical), and 203$\times$ (Speech), and by five orders of magnitude for NLP's file-level reads.
\end{insight}
 
\begin{insight}\label{ins:workers}
Decoding is the costliest stage to add in Vision and Medical and cuts throughput by 81--97\% relative to storage-only reads (97\% for Medical). Correspondingly, the CPU workers needed to keep the GPU busy range from none (NLP, where the main process alone keeps the GPU busy) to 8 (Vision and Medical).
\end{insight}

\subsection{Data Format Matters When CPU-Bound}
\label{sec:formats}
We analyze the system-level impact of using different data formats and data loaders when training end-to-end. As seen in \autoref{tab:combined-metrics} for the Vision domain, the thread-based FFCV dataloader using the Beton format keeps the GPU busiest, with a mean SMACT of 95.54\% while using 24.44\,GiB of host memory. The training throughput is also the best at 3,039 samples/s. DALI, which uses GPU for preprocessing, shows the lowest CPU utilization (mean 17.27\% of all cores) and the lowest host memory usage (13.59\,GiB). The rest of the formats (JPEG, PNG, WebDataset) use the PyTorch dataloader and show comparable memory footprint (62.25--68.27\,GiB). The PNG format has a significantly slower throughput (1,547 samples/s) because its decode stage sustains 5.8$\times$ lower throughput than JPEG's (2,018 vs.\ 11,681 images/s at 16 workers), which 16 workers are not able to mask, leading to data stalls (SMACT dropping to 0\%). 

The Medical domain in contrast shows an increase in throughput from 4.21 (NIfTI-GZ) to 4.74 samples/s (+13\%, and three back-to-back repetitions of each format show the same gap, +12\%) from using the uncompressed NIfTI format, along with a lower CPU usage (20.02\% vs.\ 24.23\%). Despite trading a 15.4$\times$ larger storage footprint (\autoref{tab:format_comparison}) for faster decoding, the pipeline remains Preprocess Bound (Bottleneck 3, mean SMACT 8.03\%) because decompression is only one component of the CPU cost, and the remaining preprocessing steps (orientation, resampling, ROI cropping) keep the CPU as the limiting stage even with the uncompressed format.

On the other hand, all speech and NLP formats show negligible impact of the data format. Both remain GPU Bound (Bottleneck 5) regardless of which format is loaded, so upstream format choice has no measurable effect on the end-to-end throughput.

\begin{insight}\label{ins:format}
Data format matters only when the pipeline is CPU bound. NLP and Speech throughput varies by under 1\% across formats. In Vision, PNG nearly halves throughput relative to JPEG (1,547 vs.\ 2,876 samples/s) because its 5.8$\times$ slower decoding exceeds what 16 workers can supply, while FFCV's pre-decoded \texttt{.beton} is fastest (3,039 samples/s). In Medical, uncompressed NIfTI raises throughput by 13\%, but stays Preprocess Bound.
\end{insight}

\begin{figure*}[t]
    \centering
    \begin{minipage}[t]{\columnwidth}
        \vspace{0pt}\centering
        \includegraphics[width=\linewidth]{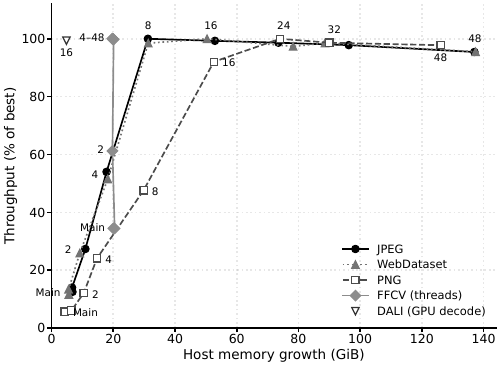}
        \caption{Throughput (\% of each loader's best) vs.\ host memory growth as DataLoader workers are added for ResNet-50, ImageNet, batch 1024. Labels give the worker count, shared by nearby points of different loaders with the same count. The unlabelled 1-worker points lie next to the main-process points, FFCV's 4 to 48 workers coincide, and DALI is shown at 16 workers. \emph{Main}: \texttt{num\_workers=0}.}
        \label{fig:multiprocss-vs-multithread}
    \end{minipage}\hfill
    \begin{minipage}[t]{\columnwidth}
        \vspace{0pt}\centering
        \includegraphics[width=\linewidth]{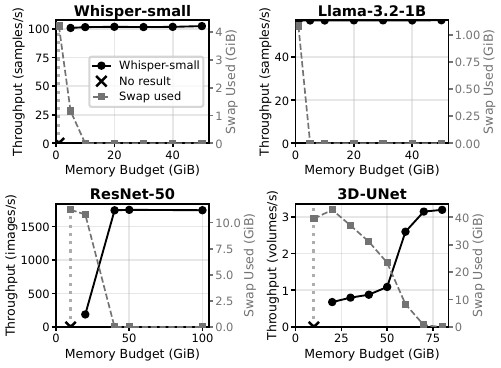}
        \caption{Domain-dependent sensitivity to constrained host memory. Throughput and swap usage vs. DRAM budget for one model per domain under constrained DRAM (\texttt{num\_workers=8}). Budgets are systemd \texttt{MemoryHigh} limits. Swap usage is on the right axis. Speech and NLP sustain full throughput down to 5\,GiB and 1\,GiB.}
        \label{fig:contrained_memory}
    \end{minipage}
\end{figure*}

\subsection{Worker Count Is a Host-Memory Decision}
\label{sec:hostmem}
\begin{figure*}[t]
    \centering
    \includegraphics[width=\linewidth]{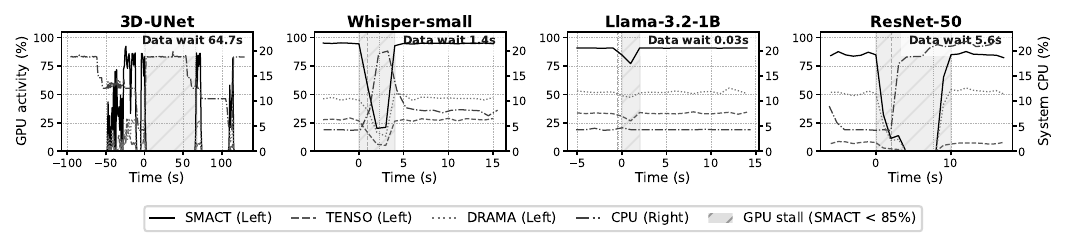}
    \caption{Epoch 1$\rightarrow$2 transition for 3D-UNet, Whisper-small, Llama-3.2-1B, and ResNet-50 ($w=8$, persistent workers), with GPU metrics on the left axis and CPU utilization on the right. Shading marks SMACT below 85\% of its pre-boundary level (1\,Hz samples), the dashed line marks the start of epoch 2, and the label gives the training loop's wait for the first batch of epoch 2.}
    \label{fig:epoch-transition}
\end{figure*}

\autoref{fig:multiprocss-vs-multithread} shows training throughput against host memory as DataLoader workers are added. For JPEG, memory in use grows roughly linearly, by about 2.7\,GiB per worker, from 6.8\,GiB above the pre-run baseline with data loading in the main process only to 137.0\,GiB with 48 workers. Throughput, however, saturates at 8 workers for JPEG and WebDataset and at 24 for PNG, and beyond that, workers only add memory. From 8 to 48 JPEG workers, host memory grows from 31 to 137\,GiB. FFCV, which runs its loader in threads (\autoref{sec:methodology}), stays at about 20\,GiB at every worker count, and DALI, which decodes on the GPU, uses 4.8\,GiB.

\begin{table*}[t]
\centering
\caption{Safetensors vs.\ PyTorch Pickle checkpoint read/write latency, mean$\pm$std [p99] over 100 runs (page cache cleared before each read), with weights and Adam moments in their native dtype (bf16 for LLMs, fp32 otherwise), as in the checkpoints written during training (\autoref{fig:degredation_throughput_plot_all}).}
\label{tab:checkpoint_format}
\footnotesize
\setlength{\tabcolsep}{2.7pt}
\begin{tabular}{@{}llll@{}cccccccc@{}}
\toprule
\textbf{Category} & \textbf{Model} & \textbf{Params} & \textbf{Ckpt. Size (GiB)} &
\multicolumn{4}{c}{\textbf{Read (s)}} & \multicolumn{4}{c}{\textbf{Write (s)}} \\
\cmidrule(lr){5-8} \cmidrule(lr){9-12}
 & & & &
\textbf{Safetensors} & \textbf{Pickle} & \textbf{Speedup} & \textbf{$\Delta$ (s)} &
\textbf{Safetensors} & \textbf{Pickle} & \textbf{Speedup} & \textbf{$\Delta$ (s)} \\
\midrule
\multirow{3}{*}{\textit{Vision}} & ResNet-50 & 26M & 0.3 & \textbf{0.17$\pm$0.004\,[0.18]} & 0.28$\pm$0.005\,[0.30] & 1.66$\times$ & 0.11 & \textbf{0.31$\pm$0.016\,[0.38]} & 0.35$\pm$0.014\,[0.38] & 1.11$\times$ & 0.03 \\
 & ViT-base & 87M & 1.0 & \textbf{0.46$\pm$0.004\,[0.47]} & 0.60$\pm$0.005\,[0.61] & 1.30$\times$ & 0.14 & 1.48$\pm$0.025\,[1.54] & \textbf{1.05$\pm$0.015\,[1.09]} & 1.41$\times$ & 0.43 \\
 & MambaVision-T & 32M & 0.4 & \textbf{0.20$\pm$0.004\,[0.21]} & 0.30$\pm$0.005\,[0.32] & 1.53$\times$ & 0.11 & \textbf{0.37$\pm$0.008\,[0.41]} & 0.41$\pm$0.005\,[0.42] & 1.10$\times$ & 0.04 \\
\midrule
\multirow{1}{*}{\textit{Speech}} & Whisper-small & 242M & 2.7 & \textbf{1.25$\pm$0.005\,[1.26]} & 1.62$\pm$0.014\,[1.65] & 1.29$\times$ & 0.36 & 3.96$\pm$0.190\,[4.46] & \textbf{3.00$\pm$0.109\,[3.29]} & 1.32$\times$ & 0.97 \\
\midrule
\multirow{4}{*}{\textit{LLM}} & Llama-3.2-1B & 1236M & 6.9 & \textbf{3.09$\pm$0.009\,[3.11]} & 5.76$\pm$0.027\,[5.83] & 1.86$\times$ & 2.67 & 13.53$\pm$0.118\,[13.77] & \textbf{10.72$\pm$0.117\,[11.01]} & 1.26$\times$ & 2.81 \\
 & Llama-3.2-3B & 3213M & 18.0 & \textbf{8.14$\pm$0.028\,[8.21]} & 14.42$\pm$0.069\,[14.53] & 1.77$\times$ & 6.28 & 34.91$\pm$0.253\,[35.39] & \textbf{27.07$\pm$0.180\,[27.48]} & 1.29$\times$ & 7.84 \\
 & Qwen3-0.6B & 596M & 3.3 & \textbf{1.50$\pm$0.008\,[1.51]} & 1.86$\pm$0.015\,[1.92] & 1.24$\times$ & 0.36 & 5.25$\pm$0.040\,[5.35] & \textbf{3.69$\pm$0.043\,[3.84]} & 1.42$\times$ & 1.56 \\
 & Qwen3-1.7B & 1721M & 9.6 & \textbf{4.36$\pm$0.030\,[4.40]} & 5.70$\pm$0.034\,[5.78] & 1.31$\times$ & 1.34 & 17.19$\pm$0.152\,[17.54] & \textbf{11.40$\pm$0.126\,[11.68]} & 1.51$\times$ & 5.78 \\
\midrule
\multirow{3}{*}{\textit{Medical}} & SegResNet & 5M & 0.05 & \textbf{0.04$\pm$0.001\,[0.04]} & 0.07$\pm$0.001\,[0.07] & 1.87$\times$ & 0.03 & \textbf{0.07$\pm$0.002\,[0.07]} & 0.07$\pm$0.001\,[0.07] & 1.10$\times$ & 0.01 \\
 & 3D-UNet & 5M & 0.05 & \textbf{0.04$\pm$0.002\,[0.04]} & 0.06$\pm$0.001\,[0.06] & 1.62$\times$ & 0.02 & \textbf{0.06$\pm$0.002\,[0.07]} & 0.07$\pm$0.003\,[0.08] & 1.11$\times$ & 0.01 \\
 & SwinUNETR & 62M & 0.7 & \textbf{0.34$\pm$0.003\,[0.34]} & 0.46$\pm$0.006\,[0.48] & 1.35$\times$ & 0.12 & \textbf{0.68$\pm$0.019\,[0.74]} & 0.79$\pm$0.015\,[0.84] & 1.16$\times$ & 0.11 \\
\bottomrule
\end{tabular}
\end{table*}

\subsection{Memory Needs Vary by Domain}
\label{sec:memory}
We now evaluate the training throughput under constrained host memory (DRAM) budgets across the pipelines to investigate their sensitivity to host memory size.
For this experiment,
we use Linux cgroups to limit the memory
and increase the swapfile from 10\,GiB to 135\,GiB 
to allow for the possible larger writes to the swap space due to the constrained memory.

\autoref{fig:contrained_memory} shows that both Whisper-small (Speech) and Llama-3.2-1B operate well under very low DRAM budgets (down to 5\,GiB and 1\,GiB, respectively). Below their working set, cold pages move to swap (1.2\,GiB for Whisper-small at 5\,GiB, 1.1\,GiB for Llama-3.2-1B at 1\,GiB) without slowing down training. In contrast, ResNet-50 (Vision) thrashes at 20\,GiB but reaches full throughput at 40\,GiB, while 3D-UNet (Medical) requires a DRAM size above 60\,GiB. 

\begin{insight}\label{ins:memory}
Host memory needs depend on both the dataloader and the domain. PyTorch's multiprocessing loader grows host memory by 126--137\,GiB at 48 workers, and at 16 workers by 2.5--2.7$\times$ as much as FFCV's thread-based loader (50.4--52.9 vs.\ 19.9\,GiB, \autoref{sec:hostmem}). Speech and NLP sustain full throughput under 5\,GiB and 1\,GiB host-memory limits, whereas Vision needs at most 40\,GiB and Medical over 60\,GiB.
\end{insight}

\subsection{Epoch Boundaries Cause Data Stalls}
\label{sec:epoch-trans}
The boundary between two epochs is also a point of particular interest, because the PyTorch DataLoader starts preparing an epoch's batches only when that epoch's iterator is created, so the dataloader cannot prepare the next epoch's first batch in advance, creating a synchronous pipeline barrier. \autoref{fig:epoch-transition} shows this transition for a model in each domain. All four traced pipelines stall at the boundary, where GPU utilization falls, as preprocessing for the new epoch only begins once the transition starts and the prior epoch's prefetch queue has drained. We measure the stall as the gap between the end of the last batch of one epoch and the start of the first batch of the next. The training loop waits 64.7\,s for 3D-UNet, 5.6\,s for ResNet-50, 1.4\,s for Whisper-small, and 0.03\,s for Llama-3.2-1B at the first transition, and 63.5, 5.2, 1.3, and 0.02\,s at the second. The wait grows with the time a worker needs to prepare one batch, which depends on the batch size as well as the per-sample cost.

3D-UNet is CPU bound (Bottlenecks 2 and 3) throughout. A worker needs a minute to prepare a batch of 32 volumes, and the loop waits up to 37\,s between batches inside an epoch, so its boundary stall is a longer instance of a stall it pays within every epoch.

Global shuffling also introduces I/O inefficiency once a text dataset exceeds host DRAM. Individual samples in the NLP domain are often only a few hundred bytes (\autoref{tab:format_comparison}), far below the readahead window the kernel reads on each page fault (256 KiB on Machine 1). A shuffled \texttt{MapDataset} reads them at random from memory-mapped Arrow files, so most of each window is not used. On the C4 dataset, this produces a 138$\times$ read amplification over three epochs versus shuffling disabled. This does not slow end-to-end training, but it wastes storage bandwidth. An \texttt{IterableDataset} with a DRAM shuffle buffer avoids this entirely (1.0$\times$ amplification), at the cost of only pseudo-random shuffling within a host DRAM buffer.

\begin{insight}\label{ins:epoch}
Every pipeline we traced stalls at epoch boundaries, for 0.03\,s (Llama-3.2-1B) to 65\,s (3D-UNet), growing with the time needed to prepare one batch, because no traced dataloader prefetches across epochs. Once a text dataset exceeds host RAM, shuffled map-style access also amplifies storage reads 138$\times$, which an iterable dataset avoids (1.0$\times$).
\end{insight}

\section{CHECKPOINTING CHARACTERIZATION}
\label{sec:checkpointing}
We now characterize reading and writing of checkpoint state
and evaluate how checkpointing affects the end-to-end training throughput across our training pipelines, using the checkpoint contents, formats, and modes described in \autoref{sec:background:checkpointing}. As seen in \autoref{tab:checkpoint_format}, the largest checkpoint in our evaluation is Llama-3.2-3B at 18.0\,GiB and the smallest are SegResNet and 3D-UNet at 0.05\,GiB.

\begin{figure*}[t]
    \centering
    \includegraphics[width=\textwidth]{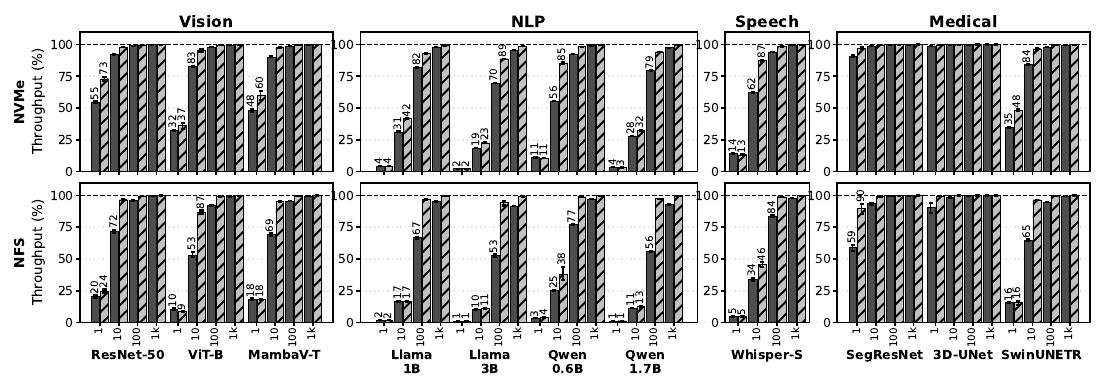}
    \caption{Throughput with periodic checkpointing, as \% of the same training without saves, for checkpoint intervals of 1 to 1000 steps (dark bars show synchronous \texttt{torch.save} with \texttt{fsync}, hatched bars asynchronous PyTorch DCP) on local NVMe (Machine~1, top) and NFS (Machine~2, bottom), as the mean of 3 runs with min--max whiskers. Bars below 90\% show their value.}
    \label{fig:degredation_throughput_plot_all}
\end{figure*}

\subsection{Checkpoint Formats Trade Reads for Writes}
We compare Pickle and Safetensors (\autoref{sec:background:checkpointing}). In \autoref{tab:checkpoint_format} we show the checkpoint read/write latency from/to disk and accelerator HBM. The absolute time difference is largest for the NLP/LLM checkpoints. Safetensors reads faster for all 11 models, by 1.24--1.87$\times$, saving up to 6.28\,s on the largest checkpoint (Llama-3.2-3B). Writes depend on checkpoint size. Pickle writes the checkpoints of about 1\,GiB or more (the smallest is ViT-base at 0.97\,GiB) 1.26--1.51$\times$ faster (up to 7.84\,s for Llama-3.2-3B), while Safetensors writes the smaller checkpoints slightly faster (1.10--1.16$\times$).

\begin{insight}\label{ins:ckpt-format}
For checkpoints of about 1\,GiB or more, no format is fastest at both reads and writes, and below that, Safetensors is faster at both. Safetensors reads every checkpoint 1.24--1.87$\times$ faster, while Pickle writes checkpoints of about 1\,GiB and larger 1.26--1.51$\times$ faster. Choose the format by whether reads or writes are on the critical path.
\end{insight}

\subsection{Async Checkpointing Pays Off When Writes Fit CPU DRAM Budget Between Saves}
\label{sec:ckpt-freq}
We measure what checkpointing costs end-to-end training on local NVMe (Machine~1) and on NFS (Machine~2), with synchronous and asynchronous checkpointing (\autoref{sec:background:checkpointing}, \autoref{fig:ckpt_timeline}). For each model we time every training step and how long each checkpoint blocks training. Synchronous saves are \texttt{torch.save} followed by \texttt{fsync}. Asynchronous saves use PyTorch DCP~\cite{pytorch_dcp} and wait for the previous write to finish before starting the next, as a training loop must in order to bound host memory. From these times we derive throughput for checkpoint intervals of 1 to 1000 steps (\autoref{fig:degredation_throughput_plot_all}, 3 runs per model, 20 warm-up steps excluded).

The cost grows with checkpoint size and frequency. A synchronous save takes 28.6\,s for Llama-3.2-3B (18\,GiB), 3.5\,s for Whisper-small, and 0.1\,s for SegResNet. Checkpointing Llama-3.2-3B synchronously every step, every 10th, and every 100th step leaves 2\%, 19\%, and 70\% of its throughput, while SegResNet and 3D-UNet keep at least 91\% even when checkpointed every step. Asynchronous checkpointing blocks training only while it copies the state to host memory (8.4\,s instead of 28.6\,s for Llama-3.2-3B), but its background write takes as long as the write portion of a synchronous save and must finish before the next checkpoint starts. It pays off when the interval is longer than the write, about 32 steps for Llama-3.2-3B. Every 10th step it keeps 23\% of throughput, barely more than synchronous checkpointing, but every 100th step 89\% instead of 70\%.

On NFS, a synchronous save takes 2.0--7.7$\times$ longer (57.7\,s for Llama-3.2-3B, 0.5\,s for SegResNet), so checkpointing every step leaves SegResNet 59\% instead of 91\% of its throughput, and asynchronous checkpointing pays off from about 86 steps for Llama-3.2-3B. Not bounding the writes trades host DRAM for throughput. Each checkpoint in flight holds a host copy of the state (18\,GiB for Llama-3.2-3B), and training is not blocked as long as DRAM can hold these copies. When checkpoints are produced faster than they are written, however, the copies accumulate until DRAM runs out. Skipping checkpoints instead keeps training unblocked but persists at most one checkpoint per write time. Frequent checkpointing also wears storage. Checkpointed every 100th step, Llama-3.2-3B writes 18--23\,TB per day. Each save writes exactly the checkpoint's size, and the SSD's write amplification is 1.01. Every 100th step thus amounts to two to three full drive writes per day of our 7.68\,TB NVMe SSD, above its rated endurance of 14,016\,TB over five years (one drive write per day)~\cite{samsung_pm9a3_2022}, while every 1000th step writes 2.4--2.5\,TB per day.

\begin{insight}\label{ins:ckpt-freq}
Checkpointing every 1000 steps costs every model less than 5\% of its throughput on NVMe and 9\% on NFS (at most 1.3\% with asynchronous checkpointing). More frequent checkpointing hurts large checkpoints. Llama-3.2-3B keeps 2\% of its throughput when checkpointed every step, while the 0.05\,GiB SegResNet and 3D-UNet checkpoints cost at most 9\% on NVMe (41\% on NFS). Asynchronous checkpointing pays off only once the interval exceeds the time of the background write (about 32 steps for Llama-3.2-3B on NVMe, 86 on NFS).
\end{insight}

\section{Design Implications}
\label{sec:discussion}

\autoref{tab:recommendations} summarizes our recommendations, the research questions they answer (\Cref{sec:methodology}), and the insights (\Cref{sec:data_loading,sec:checkpointing}) that support each one. We now delve into each recommendation.
 
\begin{table}[t]
\centering
\caption{Design recommendations, the research questions they answer, and the insights behind them.}
\label{tab:recommendations}
\footnotesize
\setlength{\tabcolsep}{3pt}
\renewcommand{\arraystretch}{1.0}
\begin{tabularx}{\columnwidth}{@{}>{\raggedright\arraybackslash}X l c l@{}}
\toprule
\textbf{Recommendation} & \textbf{Applies to} & \textbf{RQ} & \textbf{Insights} \\
\midrule
Provision for the domain-specific bottleneck & All domains & 1--3 & \ref{ins:storage}, \ref{ins:workers}, \ref{ins:memory} \\
Accelerate Medical decode, Vision preprocessing & Vision, Medical & 1, 3 & \ref{ins:nlp-gpu}, \ref{ins:decode}, \ref{ins:format} \\
Tune worker count and host memory together & Vision & 2, 3 & \ref{ins:workers}, \ref{ins:memory} \\
Use iterable datasets beyond DRAM        & Large datasets  & 4    & \ref{ins:epoch} \\
Prefetch across epoch boundaries        & All domains     & 4    & \ref{ins:epoch} \\
Match checkpointing to model size and DRAM & All models & 5 & \ref{ins:ckpt-format}, \ref{ins:ckpt-freq} \\
\bottomrule
\end{tabularx}
\end{table}
\boldfirst{Provision for the domain-specific data pipeline bottleneck.}
Vision and Medical pipelines demand high CPU compute provisioning and moderate DRAM to sustain heavy data preprocessing. By contrast, NLP pipelines use about 4\% of the host CPU (vs.\ 34--38\% for Vision with the PyTorch loader, \autoref{tab:combined-metrics}) and train at peak throughput under a 1\,GiB host memory limit (\autoref{fig:contrained_memory}). 
Furthermore, storage I/O accounted for at most 7\% of per-batch latency across all evaluated domains, and NVMe Gen 4 bandwidth exceeded end-to-end throughput by 30--203$\times$ outside NLP, and by far more for NLP.
Therefore, do not over-invest in higher generation storage bandwidth at the expense of compute for training systems.
Given the increasing DRAM and storage prices,
these insights can help with building/using cheaper servers for model training
or motivate workload collocation \cite{robroek_analysis_2024, strati2024orion} and memory sharing methods for unused and stranded CPU memory \cite{pond, 3489146.3489204}.

\boldfirst{Prioritize decode acceleration in Medical Imaging and preprocessing acceleration in Vision, not in NLP or Speech.}
Decoding consumed up to 83\% of per-batch latency for 3D-UNet and 34--60\% for other Medical/Vision models, and in Vision preprocessing took a further 42--45\% (cf. \autoref{ins:decode}). By contrast, NLP models spent over 96\% of per-batch time in GPU compute,
and file format made no measurable difference for NLP or Speech. Avoid reformatting or re-encoding data to improve training throughput in domains that are already compute-bound.

\boldfirst{Tune dataloader worker count and host memory to pipeline throughput.}
Worker count, prefetch factor, batch size, and sample dtype together set the host memory footprint of prefetched batches. With PyTorch's multiprocessing DataLoader, host memory grew by about 2.7\,GiB per worker for ResNet-50, reaching 137\,GiB at 48 workers (\autoref{fig:multiprocss-vs-multithread}). FFCV and DALI showed less memory growth at 16 workers (19.9 and 4.8\,GiB, respectively, vs.\ 50.4--52.9 GiB for PyTorch). Select the smallest worker count that reaches the throughput plateau and tune it jointly with batch size and prefetch depth. Compare dataloaders on throughput and host memory consumption together as similar training throughput can require substantially different resources.

\boldfirst{Use streaming/iterable datasets instead of shuffled map-style datasets once data exceeds host DRAM.}
For datasets larger than host memory, prefer an IterableDataset with an in-memory shuffle buffer over a shuffled MapDataset. A shuffled MapDataset on C4 produced 138$\times$ read amplification over three epochs due to sub-page-size random reads, versus 1.0$\times$ for an iterable dataset.

\boldfirst{Prefetch across epoch boundaries when the pipeline throughput drop matters.}
Every domain we traced showed a GPU utilization drop at the epoch boundary with stalls ranging from 0.03\,s (Llama-3.2-1B) to 65\,s (3D-UNet), growing with per-batch preparation time. Dataloaders should therefore begin preparing the first batches of the next epoch before the current epoch's final batch completes instead of treating epoch transitions as a synchronous barrier in case they matter for pipeline throughput.

\boldfirst{Match checkpoint format, frequency, and medium to model size and available DRAM.}
Scale checkpoint frequency to model size.
The smallest checkpoints (SegResNet \& 3D-UNet, 0.05\,GiB) can be written every step at about 9\% throughput cost on local NVMe (41\% on NFS) while the NLP models need intervals of about 100 steps with asynchronous checkpointing (89--99\% of throughput on either storage) or about 1000 steps with synchronous checkpointing (92--99\%). Use asynchronous checkpointing when the time between checkpoints exceeds the background write time (about 32 steps for Llama-3.2-3B on NVMe) and host DRAM can hold a copy of the training state. For checkpoints of about 1\,GiB or more, write with Pickle during training and re-encode to Safetensors for later reads.

\boldfirst{Research directions.} Based on our findings, future work can investigate tools for both on-prem and cloud-based deployments that jointly tune dataloader parameters (worker count, prefetch factor, and batch size) and checkpointing parameters (format, frequency, and asynchrony) for desired throughput. These tools can also guide infrastructure sizing by estimating the CPU, host DRAM, and storage bandwidth needed to sustain training throughput within memory budgets and recovery requirements. GPU-direct I/O is of interest for NLP loading and checkpointing as a way to reduce host DRAM usage and CPU staging overhead. For processing-heavy Vision and Medical Imaging pipelines, future work should evaluate GPU decoding and preprocessing and the promise of fusing operations in GPU kernels to speed up CPU-heavy preprocessing.

\section{Related Work}
\label{sec:related}

We categorize related work into three categories:
prior data-stall analysis and benchmarking studies,
systems that target specific dataloading bottlenecks,
and work on checkpointing performance.

\boldfirst{Data-stall analysis and benchmarking.} Mohan et al.~\cite{mohan_analyzing_2021} formalized host-side data stalls and introduced CoorDL to mitigate them via coordinated caching and prefetching. tf.data~\cite{murray_tfdata_2021} reported that data ingestion accounts for 30\% of total compute time across a production fleet, and Isenko et al.~\cite{isenko_where_2022} showed that preprocessing introduces further bottlenecks in tf.data pipelines. MLPerf Training~\cite{mattson_mlperf_2020}, MLPerf Storage~\cite{Balmau22}, and DLIO~\cite{devarajan_dlio_2021} each cover only part of the pipeline we characterize. The former abstracts away per-stage analysis in favor of time-to-accuracy, while the latter two replace GPU compute with a sleep call. We revisit data pipelines end-to-end on current hardware and models, across more domains, formats, and dataloaders.

\boldfirst{Dataloader systems.} Several recent systems target specific dataloading stages we characterize. SPDL~\cite{hira_scalable_2025} is a thread-based Python dataloader evaluated under free-threaded Python 3.13t. TensorSocket~\cite{robroek_tensorsocket_2025} shares a single dataloader across concurrent training processes to avoid duplicate data preparation. cedar~\cite{cedarVLDB24} optimizes the input pipeline as a whole, FastFlow~\cite{um_fastflow_2023} offloads preprocessing to remote CPUs, and FusionFlow~\cite{kim_fusionflow_2023} splits it between CPUs and GPUs. Seneca~\cite{desai_preparation_2025} combines cache partitioning with opportunistic sampling for multi-job workloads. MinatoLoader~\cite{nouaji_minatoloader_2025} batches from parallel fast/slow queues to reduce straggler impact. Our work is complementary. We provide the empirical, domain-specific characterization (across formats, worker counts, and memory budgets) that motivates when and where techniques like these are needed.

\boldfirst{Checkpointing systems.} CheckFreq~\cite{mohan_checkfreq_2021} and PCCheck~\cite{strati_pccheck_2025} both target write-induced stalls, the former via profiling-driven, pipelined snapshot/persist checkpointing, the latter via concurrent asynchronous checkpointing. Our checkpointing analysis (\autoref{sec:checkpointing}) is complementary, as we quantify how checkpoint format, frequency, and storage medium affect end-to-end throughput, showing when such techniques are needed.

\section{Conclusion}
\label{sec:conclusion}
We characterize the end-to-end data path of modern DL training across 11 models and four domains. 
Training pipelines are rarely storage-bound, as storage accounts for at most 7\% of per-batch latency, and bottlenecks range from GPU compute in NLP to CPU-side decoding and preprocessing in Vision and Medical Imaging. Small checkpoints can be written every step, while an 18\,GiB checkpoint needs about 100-step intervals and asynchronous writes. Other costs come from the dataloader itself, as workers beyond 8 add host memory but no increase in end-to-end throughput and every traced pipeline stalls at epoch boundaries. Optimizing modern DL data pipelines therefore requires looking beyond storage bandwidth toward the host-side software stack
and rethinking how we configure servers for these workloads.

\begin{acks}
This work was supported by the Innovation Fund Denmark.
We also thank Pioneer Centre for AI and the Danish Centre for AI Innovation for access to
DTU HPC and Gefion supercomputer.
\end{acks}

\bibliographystyle{ACM-Reference-Format}
\bibliography{bib}

\end{document}